# Damage-tolerant, laminated structural supercapacitor composites enabled by integration of carbon nanotube fibres

Moumita Rana[1†], Yunfu Ou[1,2†], Chenchen Meng,[1,3] Federico Sket[1], Carlos González[1,2], Juan J. Vilatela[1]

[1] IMDEA Materials Institute,C/Eric kandel 2, Parque de Technogetafe, 28906 Madrid, Spain

[2] E.T.S. de Ingenieros de Caminos, Universidad Politécnica de Madrid, 28040 Madrid, Spain

[3]State Key Laboratory of Chemical Engineering and School of Chemistry & Molecular Engineering, East China University of Science and Technology, 130 Meilong Road, Shanghai, 200237, P.R. China

[†] These authors contributed equally to this work

E-mail: carlosdaniel.gonzalez@imdea.org, juanjose.vilatela@imdea.org

## Abstract

A natural embodiment for multifunctional materials combining energy-storing capabilities and structural mechanical properties are layered structures, similar to both laminate structural composites and electrochemical energy storage devices. A structural composite with integrated electric double layer capacitive storage is produced by resin infusion of a lay up including woven glass fabric used as mechanical reinforcement, carbon nanotube non-woven fabrics as electrodes/current collectors and a polymer electrolyte. The energy-storing layer is patterned with holes, which after integration form resin plugs for mechanical interconnection between layers, similar to rivets. Finite element modelling is used to optimise rivet shape and areal density on interlaminar shear properties. Galvanostatic charge discharge tests during three point bending show no degradation of properties after large deflections or repeated load/unload cycling at 3.5 V.This mechanical tolerance is a consequence of the elimination of metallic current collectors and the effective integration of multifunctional materials, as observed by electron microscopy and X-ray computed tomography. In contrast, control samples with metallic current collectors, analogous to embedded devices, rapidly degrade upon repeated bending.

## 1. Introduction

The development of laminated structural composites over the last decades has made an enormous contribution to the reduction of weight of engineering components. Composites based on fibre-reinforced polymers have widespread use in transport, most notably in aerospace and increasingly in automotive and rail.[1–4] After successful optimisation for mechanical properties, commercial lamina of unidirectional fabrics of continuous fibres have near theoretical fibre volume fractions and tensile properties close to the upper limit set by the rule of mixtures. An attractive new challenge is to produce composites that can simultaneously act as structural elements and store energy.[5,6] This stems partly from the rapid pace of electrification of transport and a shift towards non-fossil fuel systems. But in addition to the ever increasing requirements in terms of energy density for electric propulsion,[7] there is also increasing power demand from the emergence of electronics inside vehicles, motivating for example, the integration of remote power sources that reduce weight at system level, for instance, through elmination of heavy conductors.[6] These trends call for the development of new multifunctional materials with both energy-storing and load-bearing capabilities. Recent reviews in the field of structural energy storage can be found elsewhere;[5,6] only a summary on the challenges and strategies for structural power composites is provided below.

A preferred architecture in structural energy storage are layered systems that resemble both structural laminates and electrochemical devices,[5] containing at least: two electrically conducting reinforcement fabrics separated by an electrically insulating layer, embedded in an ion-conducting polymer matrix. In battery systems ions are intercalated inside the graphitic material in carbon fibres (CF) or alloyed with an active phase attached to a carbon-based current collector, whereas in electric double layer capacitors (EDLC) energy is stored on the surface of the conducting material. For both systems, the matrix should combine high ionic conductivity and high Young's modulus, which are conflicting properties.

Multifunctional matrices can be produced by mixing ionic liquid (and often a lithium salt) with a regular structural epoxy an promoting their phase separation during curing, which results in a three dimensional bi-continuous matrix, that can reach an impressive combination of compressive modulus of up to 39 GPa and capacitance of up to 52 mF/g.[8] Such strategy has been successfully applied at full electrochemical cell level, in a system that combined activated carbon fiber as electrode, and a solid composite electrolyte consisting of organic liquid electrolyte $TEABF_4$ in propylene carbonate and an epoxy matrix. It led to electrode-specific capacitance of 101.6 mF/g and flexural modulus and strength of 0.3 GPa and 29.1 MPa respectively.[9] Phase-separated matrices are an elegant method to produce bi-continuous multifunctional matrices, but are challenging to produce outside a narrow thermodynamic window and difficult to achive in combination with mesoporous electrodes.

Alternatively, the problem of simultaneous interconneting plies for stress transfer and ion transport can be circumvented by patterning the multifunctinal layer through introduction of holes in the energy-storing lamina. After infusion the holes act as interconnecting points, conceptually similar to rivets, forming a macroscopic bi-continous structure only in the direction of interest, i.e. normal to the plies. This routes has been previously used for structural supercapacitors[10] and more recently structural batteries.[11]

For structural EDLCs the primary building blocks are conductive high surface area carbon-based materials.[12] CFs have exceptional mechanical properties but an intrinscially small specific surface area (SSA) of around 0.2 $m^2$/g. Early methods to increase SSA included combination with CNTs as sizing or directly grown on the CF surface, which increased capacitance from <0.2 F/g to 4 F/g.[13] Shaffer and coworkers further developed a method to grow carbon aerogels around CFs, followed by resin infusion under flexible tooling, which produced a porous matrix with high shear modulus (895 MPa) and specific capacitance of 602 mF/g.[14] Alternatively, it is possible to directly use CNTs, which even in macroscopic ensembles preserve SSA above 100 $m^2$/g. Recently, Muralidharan et al. have developed CNT reinforced structural composite supercapacitor by growing CNTs on stainless steel mesh, combined with $LiBF_4$ in $BMIBF_4$ electrolyte. The device exhibited energy density up to 3 mWh/kg along with elastic modulus over 5GPa and tensile strength greater than 85 MPa.[15]

In previous work we presented a route for structural supercapacitors using unidirectional non-woven fabrics of CNT fibres (CNTF), which have a combined specific surface area of around 250 $m^2$/g, specific capacitance from 10 to 20 F/g and high mechanical toughness (fracture energy). The CNTF fabrics can be assembled together with a polymer electrolyte to form large-area all-solid supercapacitors, stable for over 10,000 galvanostatic charge-discharge cycles.[16] These CNTF/polymer electrolyte materials where previously embedded with reinforcing fibres and a thermosetting matrix to form laminate composites. However, further studies on coupled mechanical and electrochemical studies where hampered by the large presence of defects in the structure, most notably those arising from the presence of thick metallic current collectors, which essentially act as severe delaminations.

Here, we present results on the fabrication of structural supercapacitor composites by resin infusion of glass fibre (GF) fabrics after integration of thin layers of CNTF fabric and polymer electrolyte without metallic current collectors. Coupled mechanical and electrochemical measurements are conduted under various loading conditions to asses the robustness of the multifunctional composite, showing resilience to fatigue in bending and no degradation of electrochemical properties. The multifunctional composite outperforms control samples with metallic current collectors with respect to both energy/power density and durability, highlighting the differences between multifunctional materials and multifunctional systems.

## 2. Experimental details

### *2.1 Materials*

Unidirectional CNT fibre fabrics supplied by Tortech Nanofibers. They were produced by winding multiple individual CNT fibre filaments onto a spool. Synthesis conditions were adjusted to produce few-layer multiwalled CNTs. Typical dimensions of the nonwoven CNT fibre fabrics were 8 x 19 $cm^2$, corresponding to an aerial density of ~ 1.6 $mg/cm^2$.

Plain woven glass fabrics of 22$g/m^2$ were used for mechanical reinforcement. The polymer matrix used consists of DERAKANE 8084 elastomer-modified bisphenol-A epoxy vinylester (Ashland Inc.), MEKP hardener and cobalt octoate catalyst, which is suitable for vacuum bag infusion processes and very convenient for fabrication of the multifunctional composites presented in this work[17].

### *2.2. EDLC layer assembly*

EDLC layer assembly was performed based on a method reported earlier[16,18]. Briefly, the polymer-ionic liquid membrane was prepared by dissolving PVDF-co-HFP and $PYR_{14}TFSI$ (4:6) in acetone and casting it at room temperature on a flat surface using doctor blade. Similarly, a PVA membrane was prepared by dissolving PVA into a mixture of water and ethanol (1:1) at 95 °C and casting it at room temperature. The thickness of the polymer electrolyte and PVA membranes were optimised at 60 and 15 μm, respectively. The membranes were dried at room temperature. To assemble the EDLC, the polymer electrolyte-ionic liquid membrane was sandwiched between two free-standing CNT layers, followed by two protective PVA membranes. The layers were consolidated by applying a force of 2 tons for 10 minutes, followed by lamination using a commercial lamination machine.

Hole patterns were produced by cutting EDLC layers with a ZING 16/24 laser cutting machine (Epilog Laser). The hole pattern was designed based on finite element (FE) simulations used to determine the effect of hole size, shape and number on the relative shear strength. Full details of the model will be provided in a future publication.

### *2.3. Structural supercapacitor fabrication*

To prepare the laminates, 16 layers of the plain-woven glass fabric were stacked together following the $[0^o]_{16}$ layup. The EDLC devices were integrated in the mid-plane of the laminate. To facilitate resin impregnation, two layers of a highly porous distribution media were placed on the both sides of the laminate. The laminate kit containing the supercapacitor device was enclosed within in a sealed bag and connected to a resin inlet and a vacuum outlet for infusion. The resin was first prepared by mixing Derakane resin, MEKP hardener and Cobalt octate at weight ratios of 100:1:0.2 and degassed prior to infusion. Due to the low viscosity of the resin, the full fabric impregnation process was completed in less than 5 minutes. After infusion, the composite panel with the integrated supercapacitor was left to cure for 72 hours at room temperature. Coupons were extracted from the panel using an electric saw and edges were polished by manual grinding with sand paper. The fabrication method and laminate lay-up were purposely designed to produce samples for in-situ electrochemical studies under mechanical loading, the main objective of this paper. Fabrication parameters may need to be modified for accurate assessment of the full mechanical performance of the multifunctional laminates in terms of tension, compression, shear, impact and toughness.

### 2.4. *Electrochemical and mechanical testing*

Electrochemical (EC) characterization involved performing Galvanostatic charge-discharge (CD), and electrochemical impedance spectroscopy (EIS), using a Biologic potentiostat (SP200). The capacitance ($C_{sp}$) of devices was calculated from the slope of the discharge profile:

$$C_{sp} = I/slope$$

Where $I$ is the applied current. Real energy and real power density values were calculated by integrating the discharge profile of the full devices using the following equations:

$$\mathrm{E_{real}} = \mathrm{I}\int \mathrm{Vdt}$$

$$\mathrm{P_{real}} = \frac{\mathrm{E_{real}}}{\mathrm{t_{dis}}}$$

Where $V$ is the voltage and $t_{dis}$ is the discharge time.

### 2.5. *Mechanical and electrochemical testing*

With the aim of understanding the factors limiting electrochemical performance under operation of the integrated structural supercapacitors, we selected the three-point bending test as a convenient method to carry out in-situ electrochemical measurements while subjecting the layers to bending and shear deformations.

Mechanical tests were performed on rectangular cross-section beams using a universal electromechanical testing frame (INSTRON 3384), according to ASTM D790 standard: beam dimensions of length, width and thickness were set to 250.0 mm x 40.0 mm x 2.9 mm, respectively, while the span distance between supports was 200 mm. The load-displacement curve was recorded for each of the tests with the INSTRON data acquisition system. Bending stiffness, obtained from the initial slope of the load-displacement curve, was consistent among the tests and varied only slightly between 36 N/mm and 41 N/mm.

The testing procedure consisted in three consecutive three-point bending tests with the associated electrochemical measurements. First, samples were subjected to three-point bending at a constant stroke of 2 mm/min until a mid-span deflection of 8 mm, achieved in continuous steps of 1 mm. At each step, CD and EIS measurements were carried out under displacement control. The same method was applied for specimen unloading. The static tests were followed by the application of 50 load-unload cycles with mid-span deflection of 5 mm under stroke control of 8mm/min in order to evaluate a possible electrochemical degradation of the integrated devices submitted to a simple "fatigue" cyclic loading. In this case, after each 10 bending cycles CD and EIS measurements were performed in the corresponding unloaded state. Finally, samples were subjected to monotonous loading until a clear fiber breakage sound was detected. Normally, this event occurred for around 12 mm mid-span deflection. CD and EIS measurements were performed again to ascertain possible permanent damage induced to the energy-storage elements of the composite.

### *2.6 X-ray Tomography measurements*

Prismatic samples of ≈ 45×5×3 $mm^3$ were extracted from the center of each of specimens by using a diamond wire. The internal microstructure of the composite containing EDLC was studied by X-ray Computed Tomography (XCT) using a Nanontom 160NF (GF-Phoenix). Tomograms were collected at 75 kV and 110 μA using a molibdenum target. For each tomogram, 1900 radiographs (averaging 8 consecutive radiograph and skipping 1 for the next projection) were acquired with an exposure time of 500ms. Tomogram voxel size was set to 3 μm. The tomograms were then reconstructed using an algorithm based on the filtered back-projection procedure for Feldkamp cone beam geometry. The acquisition time for each tomogram was around ≈ 2.4 h.

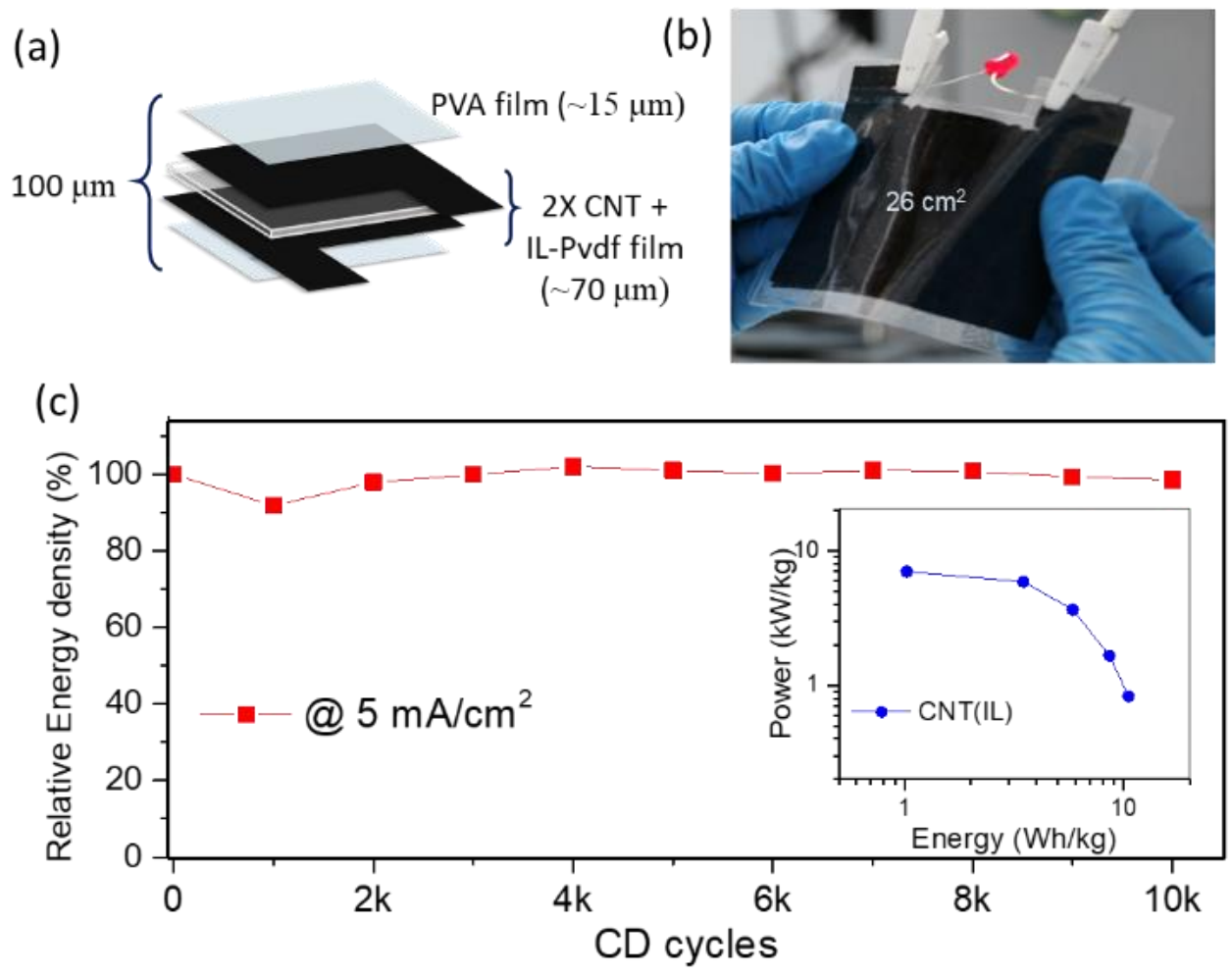

***Figure 1.*** *EDLC layer for integration in structural supercapacitor composite (a) Schematic of the EDLC composition. (b) Digital image of a thin, flexible EDLC, lighting a red LED. (e) Relative energy density of the EDLC upon galvanostatic cycling at current densities of 5* $mA/cm^2$ *for 10,000 galvanostatic cycles. Inset shows ragone plot.*

## 3. Results and discussion

The structural supercapacitors produced in this work consist of laminates of woven glass fibre fabric containing an integrated layer that can store energy by acting as an EDLC. The first step is producing the EDLC by simply pressing two CNT fibre fabrics separated by a polymer electrolyte membrane comprised of ionic liquid and polymer electrolyte (PVDF-co-HFP). Because of its high ionic liquid content, the membrane can penetrate into the porous CNT fibre fabrics under pressure even at room-temperature. Two additional thin polymer layers were used as protector and insulating layer over the CNTF fabrics. **Figure 1a-c** presents the schematic of an EDLC construction, and examples of devices and their electrochemical properties.[19]

The resulting EDLC has a thickness of ~100 µm, with ~60 µm coming from the polymer membrane that also acts as separator (**Figure S1a**). Such current collector-free, flexible configuration leads to EDLC with ~ 624 mF/g capacitance, 656 W/kg power density, 1.12 Wh/kg energy density, as well as 98.6 % retention of capacitance after 10000 charge-discharge cycles at high current density of 5 mA/cm$^2$. Very importantly, the CNT fibre fabric acts both as electrode and as current collector. In this configuration without metallic elements and a reduced polymer electrolyte membrane thickness relative to previous work, a full structural capacitor composite can reach 246 W/kg power density and 0.42 Wh/kg energy density (**Figure S2**).

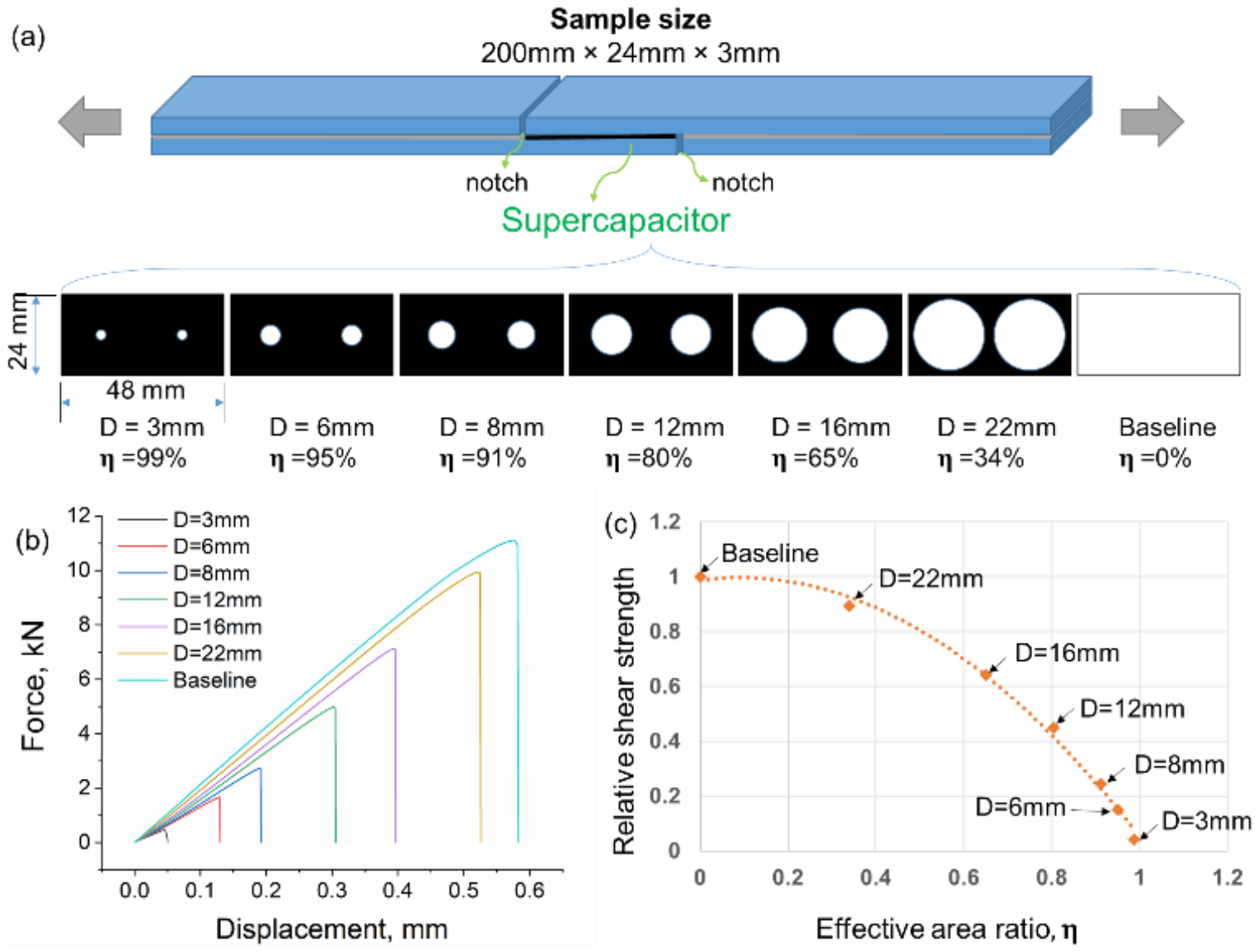

*Figure 2. Optimisation of hole pattern structure for multifunctional performance. (a) Schematic representation of the sample configurations and dimension for single lap-shear simulationt; (b) Representative force-displacement curves of lap-shear specimens with different effective area and (c) relative shear strength for different interleaf area.*

From a mechanical perspective, the introduction of a flexible and deformable film inside the composite laminate can induce strong displacement distortions because of the presence of non-connected plies. Shear connectors can alleviate such distortions caused by the presence of the supercapacitors. The strategy consists in using EDLC layers with a pattern of holes as supercapacitor interleaves. Upon infiltration, resin fill the empty spaces created by the holes interconnecting the adjacent laminate layers with resin structural rivets enabling an effective shear transmission between the plies[10].

There are almost endless combinations of different size, shape and number of holes, limited only by the precision of the patterning equipment (see examples in **Figure 2a** and **S3**). As a rational strategy to select a relevant combination of parameters close to an optimum, we performed FEA (finite element analysis) simulation on double slotted single-lap shear specimens on 6 model samples with different effective area. The diameters of rivets simulated were 3, 6, 8 and 16 mm. **Figure 2b** shows the computed force-displacement curves of the double slotted lap-shear specimens with different effective areas. The relative shear stiffness, initial slope of the curve, and shear strength, maximum of the curve, increased with increasing diameter of reinforcing rivets. The dependence of relative shear strength (normalized by that of the fully bonded specimens) on effective area ratio is proved to be slightly nonlinear, (**Figure 2c**). The modelling details, including geometry and input parameters can be found in ESI.

***Table T1***. *List of samples subjected to mechanical three point bending tests and electrochemical measurements.*

| **Samples** | $A_{EDLC}$ | **Current collector** |
|---|---|---|
| **CNTCC-100** | 100% | no |
| **CNTCC-83** | 83% | no |
| **AlCC-100** | 100% | Al |
| **AlCC-83** | 83% | Al |

Based on simulation results, we chose equally spaced, circular rivets of 0.8cm in diameter, producing an array with ~83% areal fraction of supercapacitor material (containing ~17% of rivets for shear transmission). This architecture provides relatively large power/energy density while retaining around 20% the shear strength relative to the fully connected laminate strength (**Figure 2c**). Laminates containing supercapacitors without rivets were also produced for comparison.

To understand the role of current collector and the riveting area of the EDLC on the electrochemical performance under the action of mechanical load, four types of samples were produced, as summarised in **Table1**. The laminates consist of composite samples with the different combinations of supercapacitors: with and without rivets, and with either aluminium foil as current collector or with the CNT fibre material itself as current collector.

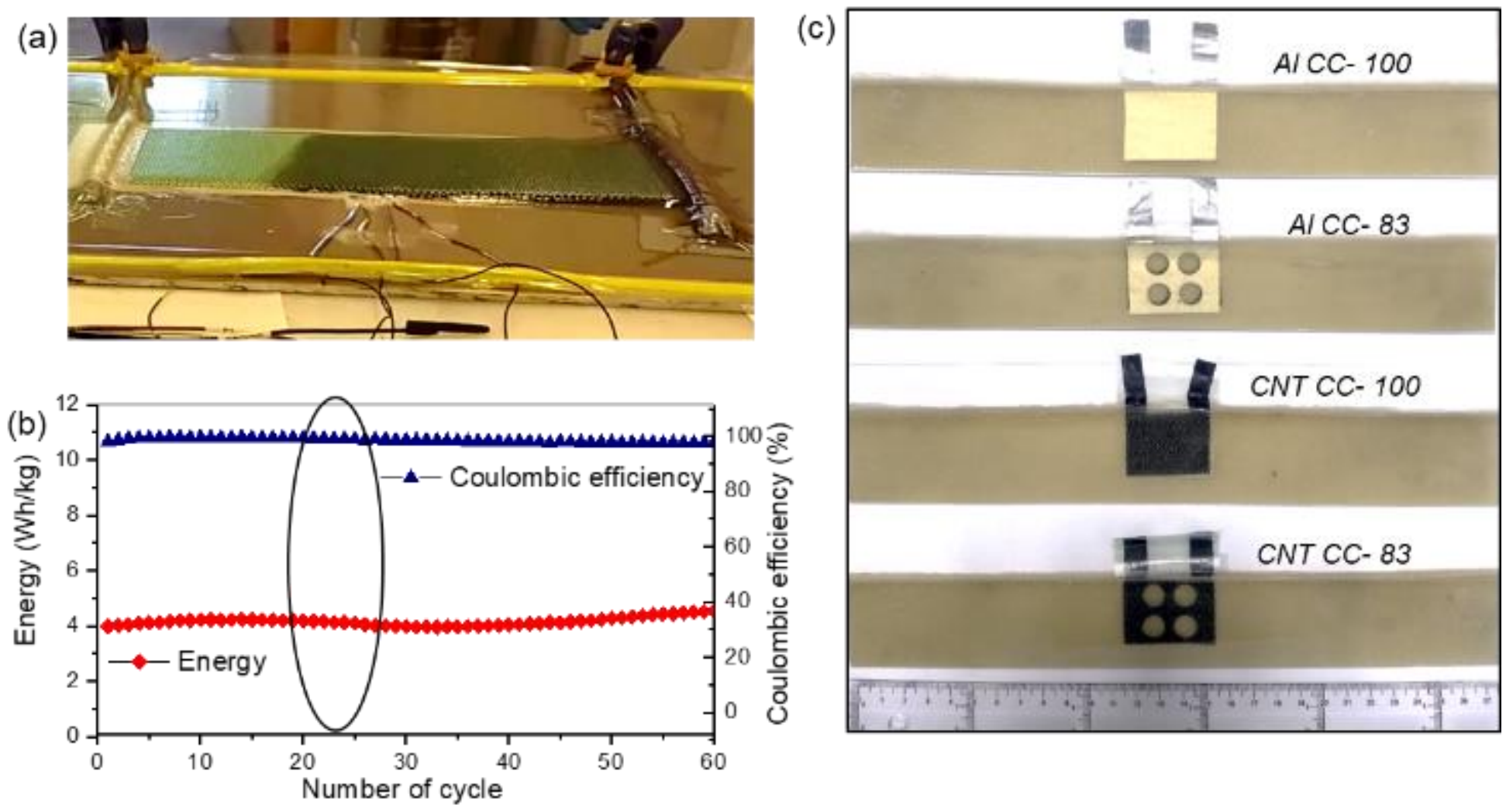

***Figure 3**. (a) Digital image of the in situ infusion process. (b) Plot of Energy and coulombic efficiency over CD cycles during the in-situ infusion experiment. (c) Digital image of the EDLC after curing of resin. The four laminated samples named as device Al CC-100, Al CC-83, CNT CC-100 and CNT CC-83 as tabulated in **Table 1**.*

After its assembly, the EDLC layer was laid up between GF plies and the whole lay-up vacuum infused with the Derakane thermosetting polymer. In addition to its impact on mechanical and electrochemical properties, the choice of matrix and polymer must be taken into account to ensure adequate compatibility between the components. CD and EIS measurements can be conveniently performed in situ during resin infusion to assess compatibility and other possible effects during fabrication (**Figure 3a).** Indeed, when using a free-standing EDLC without any external current collector or insulation, the resin dissolved IL from the electrolyte membrane and caused electrochemical failure after composite fabrication, as shown in **Figure S4.** Replacing the IL-based electrolyte with an aqueous one is a possible simple solution (**Figure S5**). However, the higher operating voltage of IL and the corresponding higher power density of the devices, makes it is of interest to develop a general method for integration of IL-based ELDC layers in structural supercapacitor composites.

Our strategy is to introduce a thin PVA membrane (~15um in thickness) between the EDLC and the GF plies. This layer can also be conveniently used as electrical insulation when using CF fabrics. **Figure 3b** shows the energy density and coulombic efficiency of an EDLC measured in situ during the infusion process (see highlighted area). No change in energy density or coulombic efficiency was observed, which indicates a successful infusion process without influence on the electrochemical performance of the composite. **Figure 3c** shows the laminated EDLC composites after the infusion and curing processes.

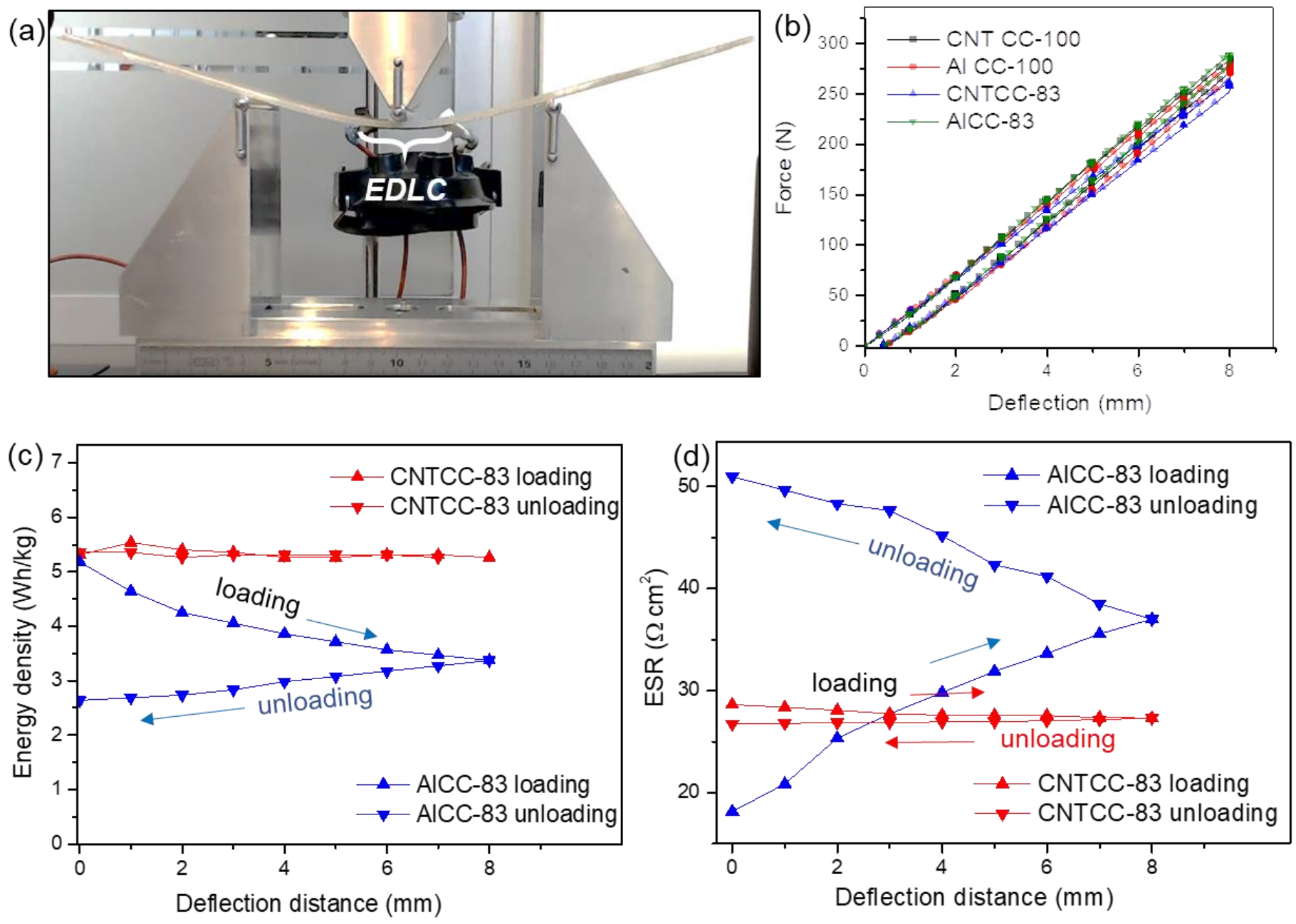

***Figure 4****. Electrochemical measurements during three-point bending (a) Digital image of the setup used for in situ measurements, with the sample under a deflection of 12 mm. (b) Load-displacement curves of the three-point bending specimens.CD and EIS measurements are performed during progressive deflection. (c) Plot of energy density against deflection for the different types of samples (d) Plot of energy density against deflection. Both (c) and (d) show no major changes for the CNTF samples, and rapid degradation for those with Al current collector.*

After curing, these four devices were subjected to electrochemical measurements during quasi-static three-point bending tests (**Figure 4a)**. The sequential testing protocol consisted in: i) in-situ CD and EIS during progressive 1mm deflection up to 8mm deflection ii) 50 load/unload cycles to 5mm deflection and CD and EIS every 10 cycles, iii) loading to sufficiently large deflection (12 mm deflection) until hearing acoustic emissions indicating a possible onset of fibre fracture. This latter step was followed by CD and EIS in the unloaded state. All charge-discharge profiles and electrochemical impedance spectroscopy data are included in Supplementary Information (**Figure S6 – S10)**.

The changes in electrochemical properties throughout progressive deflection of patterned samples with and without Al-CC are summarised in **Figure 4**. Their mechanical behaviour was almost identical, with a nearly linear and elastic behaviour and minor evidences of plastic deformation or damage. Possible differences in bending stiffness can be attributed to sample preparation although they have no significant effect on the interpretation of electrochemical data recorded. Throughout the load and unload process the EDLCs without CC show a negligible change (<1.8%) in energy density; in contrast, the EDLCs with Al-CC undergo significant loss (>30%) in energy density (**Figure 4c**). The main cause of this degradation is a continuous increase in electrical resistance, which in terms of ESR determined from EIS (**Figure 4d**) can increase by as much as 180%. For reference, the change in ESR of the multifunctional CNTF composite is a drop of 9%.

The data in **Figure 4b** also indicate that the deterioration of the Al-CC sample continues during unloading. This mechanism was further studied by performing a "fatigue" test consisting of 50 bending cycles with 5 mm mid-span deflection, with CD and EIS collected every 10 bending cycles. Again, a nearly constant bending stiffness (**Figure 5a**) suggests no major change in mechanical properties of the overall composite that could propagate to the EDLC layer. A comparison of CD cycles before and after the test shows that the CNTF sample remains unaltered, whereas the Al-CC specimen degrades from an already poorer performance (**Figure 5b, c**). As suspected, the latter undergoes continuous loss el energy density due to constant increase in ESR during repeated load/unload cycles (**Figure 5d**).

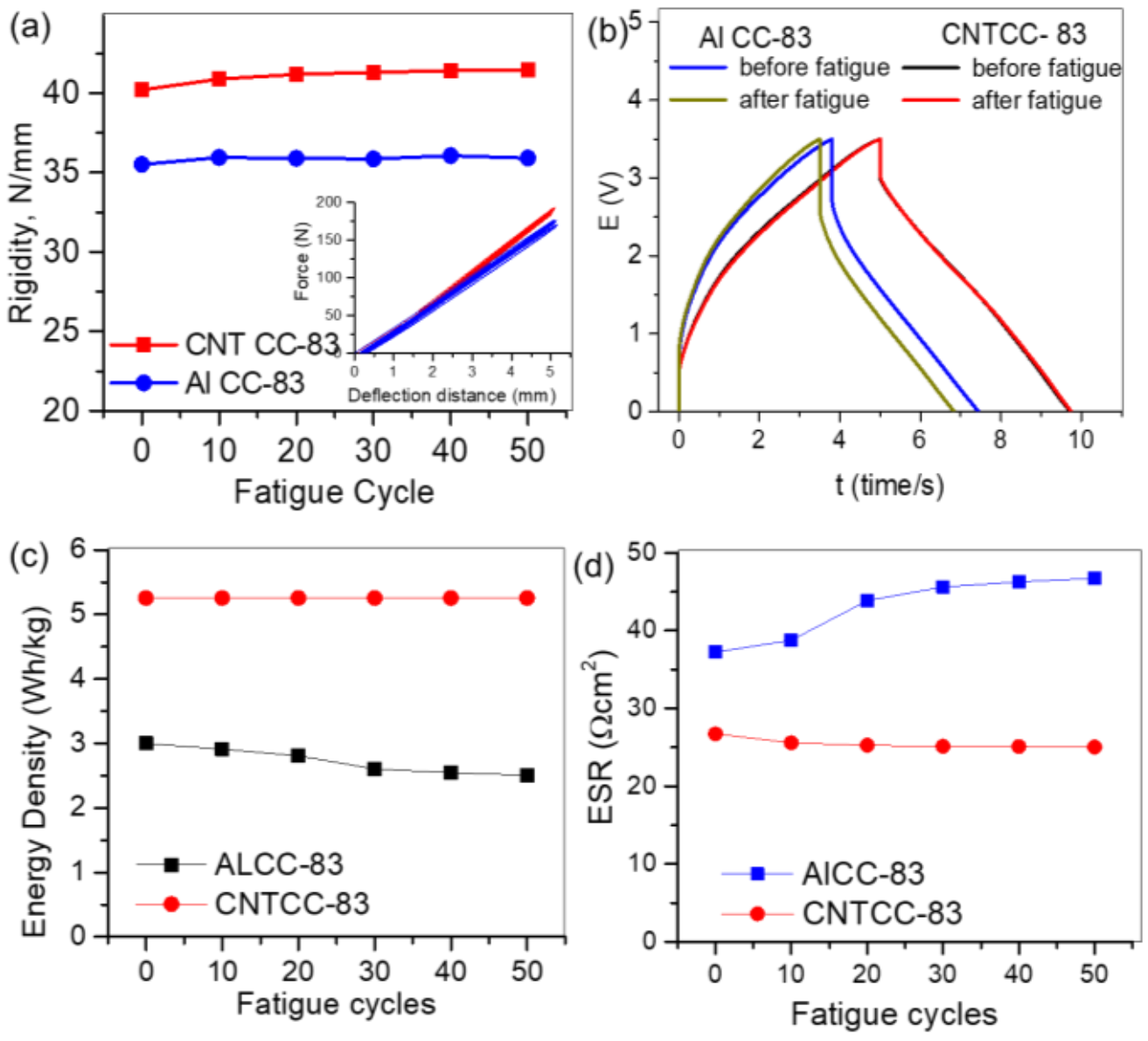

***Figure 5****. Load/unload bending cycles to 5 mid-span deflection, and in-situ electrochemical measurements. (a) No major changes in bending stiffness are observed, indicating no degradation of mechanical properties. (b) CD profiles show no degradation of samples with CNTF current collector, but poor performance for Al-CC control samples. (c) Plot of energy density against bending cycle showing progressive damage during cycling when using Al-CC. (d) ESR determined from EIS is constant for CNTF devices but increases continuously during fatigue cycling and is the main cause for electrochemical deterioration of Al-CC comparator.*

Finally, a more extreme bending test was performed by bending each laminate up to a large deflection distance of 12 mm, at a force of ~450 N (**Figure S9 and S10**). Under this large deflection, fracture of glass fibres could be heard and white crack marks in the laminate could be observed. In spite of producing a large deflection reaching the onset of GF fracture, CD and EIS profiles of the EDLCs before and after the test show that the composite can store energy and the sample with CNTF shows no electrochemical degradation. In contrast, the EDLCs with aluminium CC experienced a significant drop in the discharge time as well as increase in resistance.

We gain insight into the effect of repeated bending tests on electrochemical properties by plotting the relative change in energy density and in ESR throughout the complete testing procedure, as shown in **Figure 6**. As noted before, the samples with CNTF show no degradation of electrochemical properties, neither in continuous planar ELDC layer configuration nor with hole patterns for interplay reinforcement.

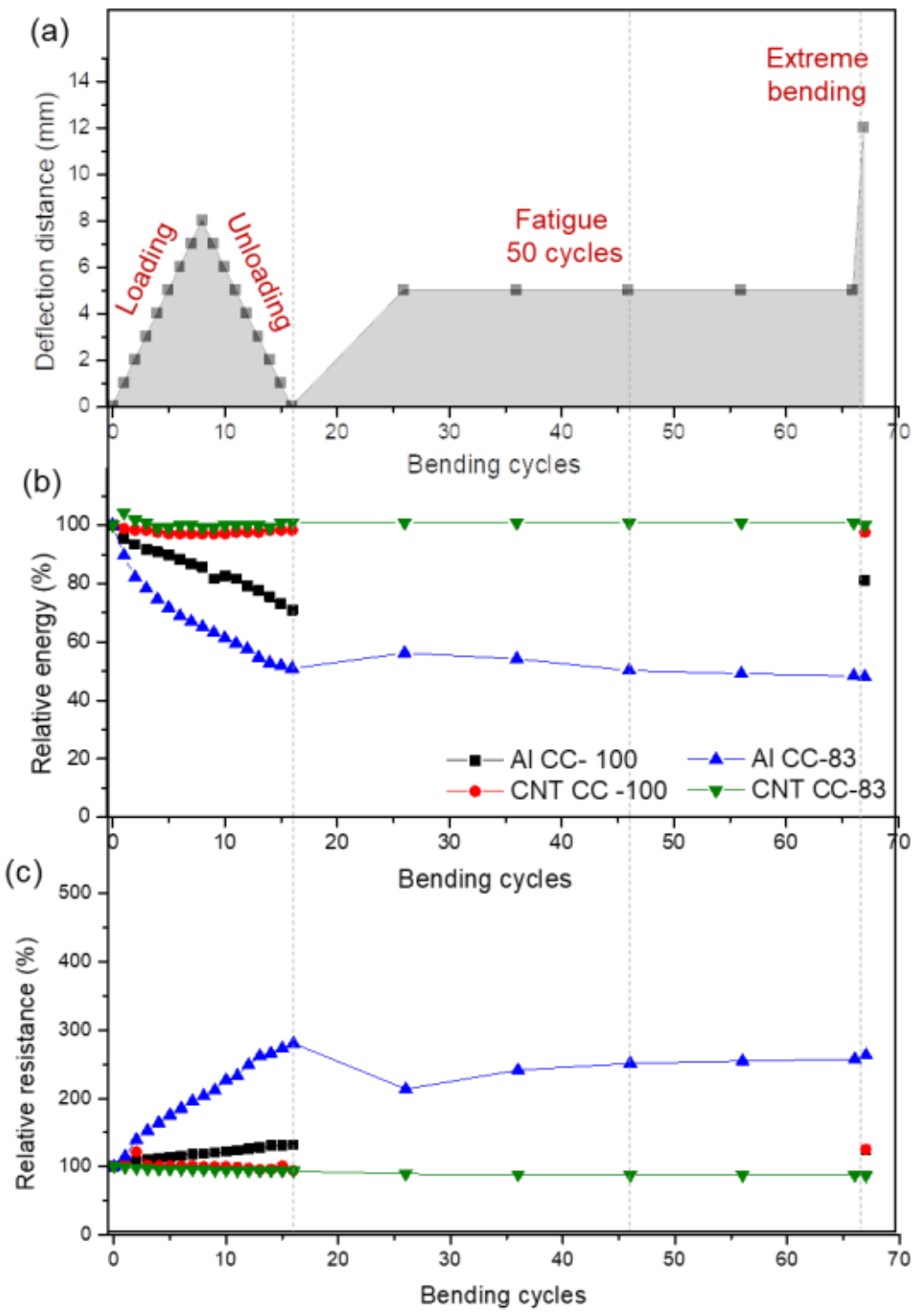

***Figure 6***. *Variation of (a) deflection length, (b) relative energy density and (c) relative resistance during in-situ electrochemical load/unload bending cycles.*

In contrast, the samples with Al-CC, analogous to embedded devices, show a loss of energy density, mainly through an increase in ESR, causing an overall 80% decrease in energy density. Since this degradation occurs continuously during both loading and unloading cycles, it is related to failure of interfaces.

To further investigate the origin of the electrical/electrochemical degradation mechanism we performed 3D X-ray tomography inspection of the central part of supercapacitor composites with and without current collector. **Figure 7** and **S11** show the reconstructed images of the cross-section of different surfaces of integrated EDLC with. The EDLC with Al CC shows prominent areas of voids between the GF layer and the EDLC (**Figure 7a**), but very importantly, there is extensive internal delamination of the ELDC. Electron micrographs confirm that there is detachment of the CNTF layer from the Al CC (**Figure 7b** and **S12**), which we identify as the origin of the increase in electrical resistance and progressive degradation of electrochemical performance in the composite through successive loading. In contrast, the devices without metallic CC adapt to the wavy shape of the GF plies (**Figure 7c, S11e-g**). There is evidence of detachment of the EDLC from the GF layer after bending, which is expected, hence the introduction of resin rivets, but the EDLC layer of this composite remains fairly intact, with no evidence of internal delamination that could increase electrical resistance or other forms of degradation.

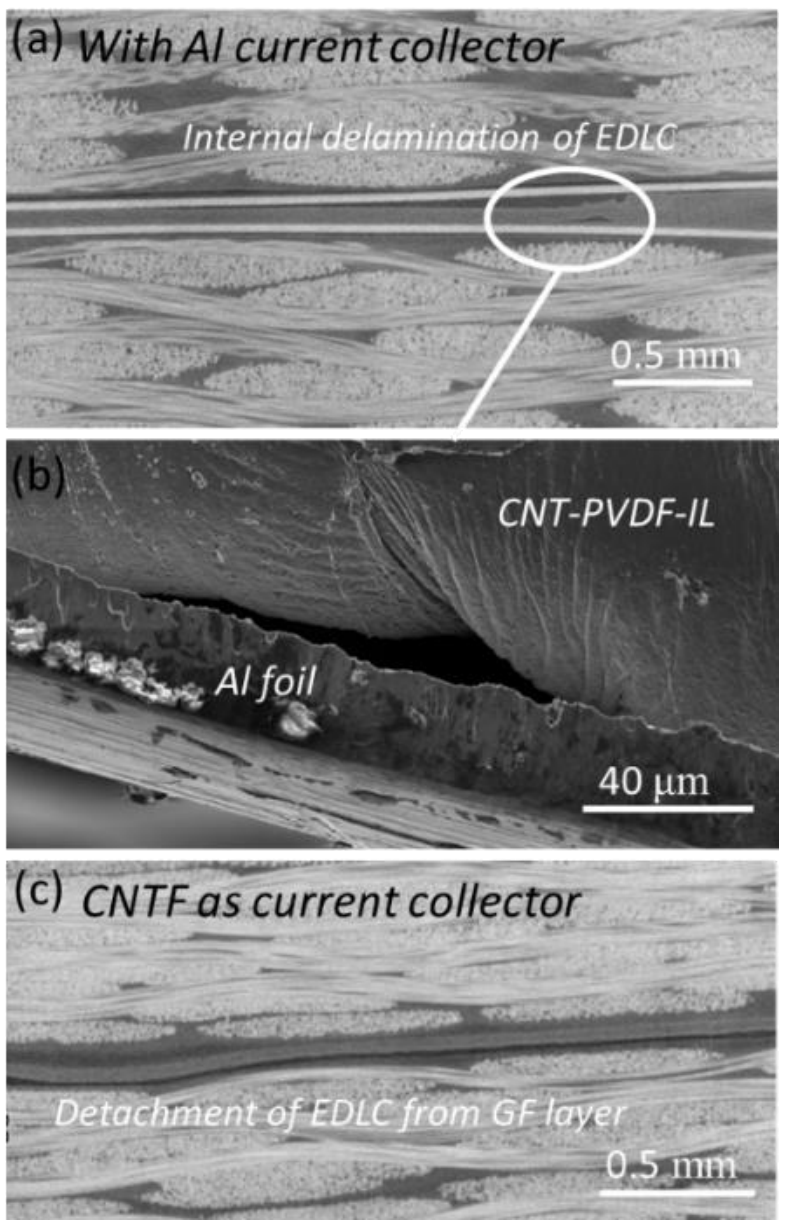

***Figure 7****. Reconstructed images of the cross-sections of composite containing EDLC (a) with Al CC and (c) without CC from 3D X-ray tomography experiments. (b) FESEM image of the EDLC with Al CC showing its detachment from the CNT-electrolyte membrane layer, similar to a buckling condition shown by the white arrow in (a).*

Our detailed analysis on the effect of mechanical deformations on the electrochemical performance of the different EDLC configurations infers that the ones with only CNT current collector are the ideal candidate for preparation of structural supercapacitor. Such superior performance of the CNT fiber fabric electrodes can be attributed to its multifunctional characteristics involving its mechanical stability, high conductivity and electrochemical activity. To demonstrate on the scalability and potential of the process, we have produced a representative large area laminate (~1000 $cm^2$), containing partial active area of EDLC component. Figure 8a shows the image of the laminate, where the effective area of individual EDLC interleaves of 24.5 $cm^2$. The EDLCs were embedded in 3 parallel layers (top, middle and bottom), where each layer contains 5 EDLCs. The connections of each EDLC electrodes were left open outside of the laminate in order to obtain necessary connections in series and/or parallel. No change was observed in the electrochemical performances of the EDLCs before and after the resin infusion process. The composite could be successfully charged up to a voltage of 30 V. The discharge profiles of the structural supercapacitor composite at charging voltages of 6, 10, 15 and 30 V are shown in Figure 8b.

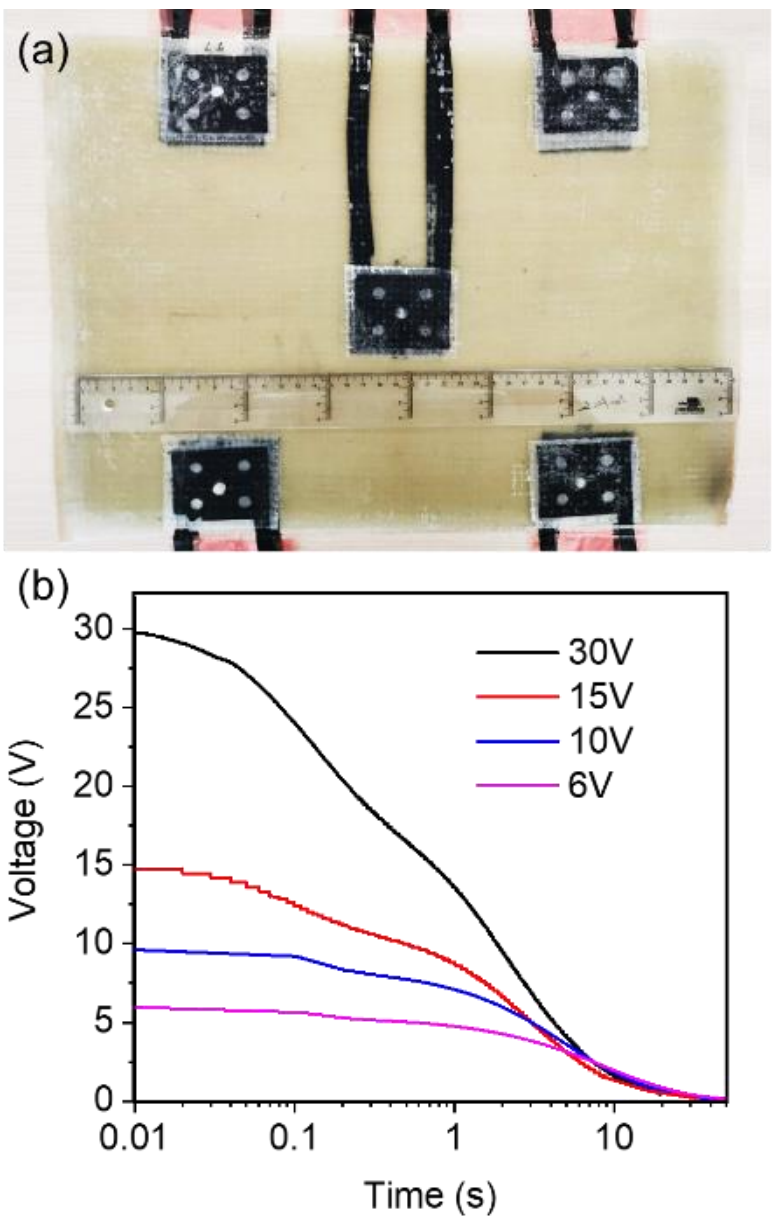

***Figure 8***. *(a) A large area structural supercapacitor containing 15 (5 x 3) EDLC interleaves (laminate area ~ 1000 cm*$^2$*). (b) Discharge profiles of the structural supercapacitor at different charging voltages.*

## 4. Conclusion

Structural supercapacitor composites were fabricated by vacuum bag resin infusion of a lay-up containing glass fibre, CNT fibre fabrics and an IL-based polymer electrolyte, without metallic current collectors. Dissolution of the polymer electrolyte by the vinyl ester resin was avoided by introducing a thin PVA layer between GF and the CNTF electrode.

Mechanical interconnection of GF plies through the EDLC can be achieved by patterning holes in the ELDC layers, which are subsequently filled with resin during infusion and form resin plugs, or rivets. With a pattern of circular holes of 8 mm diameter and a fraction of 17% holes relative to EDLC, projected power and energy densities of 172 W/kg and 0.3 Wh/kg, respectively, could be reached, combined with an estimated 20% of interlaminar shear strength relative to the fully connected composite layers.

The robustness of the structural supercapacitor composites was evaluated by a three-point bending protocol including progressive deflection, repeated load/unload cycles and deflection up to GF fracture, combined with in/ex situ electrochemical measurements. The multifunctional material developed in this work shows no appreciable degradation of its electrochemical properties, a consequence of the use of the CNT fibre fabric electrode as current collector. In contrast, control samples with Al current collectors, analogous to embedded devices, show rapid and progressive degradation of their properties, mainly caused by an increase in electrical resistance through internal delamination of the ELDC components. Further, we demonstrate the scalability of the process by producing a large area structural supercapacitor (1000 $cm^2$) for high voltage applications.

In conclusion, the use of CNT fibre current collector increases energy/power density and overall figures of merit in energy-storing structural composite. But very importantly, it makes the electrochemical elements tolerant to mechanical deformations by preserving electrical contact of electrochemically-active elements. This feature is critical for continuous operation of batteries, for example, and indicates that the multifunctional materials approach used is a key enabler for the safe operation of structural energy-storing composites.

## Acknowledgements

The authors are grateful to A. Mikhalchan for patterning EDLC samples and to Tortech Nanofibers for provision of CNT fibre fabric samples. M.R. and J.J.V. are grateful for generous financial support provided by the European Union Seventh Framework Program under grant agreement 678565 (ERC-STEM) the Clean Sky Joint Undertaking 2, Horizon 2020 under Grant Agreement Number 738085 (SORCERER), and by MINECO (RyC-2014-15115). Y. Ou appreciates the financial support from the China Scholarship Council (grant number 201606130061) and the Cost Action CA15107 (MultiComp).

**Supporting information**

# Damage-tolerant, laminated structural supercapacitor composites enabled by integration of carbon nanotube fibres

**Moumita Rana[1†], Yunfu Ou[1,2†], Chenchen Meng,[1,3] Federico Sket[1], Carlos González[1,2], Juan J. Vilatela[1]**

[1] IMDEA Materials Institute,C/Eric kandel 2, Parque de Technogetafe, 28906 Madrid, Spain
[2] E.T.S. de Ingenieros de Caminos, Universidad Politécnica de Madrid, 28040 Madrid, Spain
[3]State Key Laboratory of Chemical Engineering and School of Chemistry & Molecular Engineering, East China University of Science and Technology, 130 Meilong Road, Shanghai, 200237, P.R. China
[†] These authors contributed equally to this work

E-mail: carlosdaniel.gonzalez@imdea.org, juanjose.vilatela@imdea.org

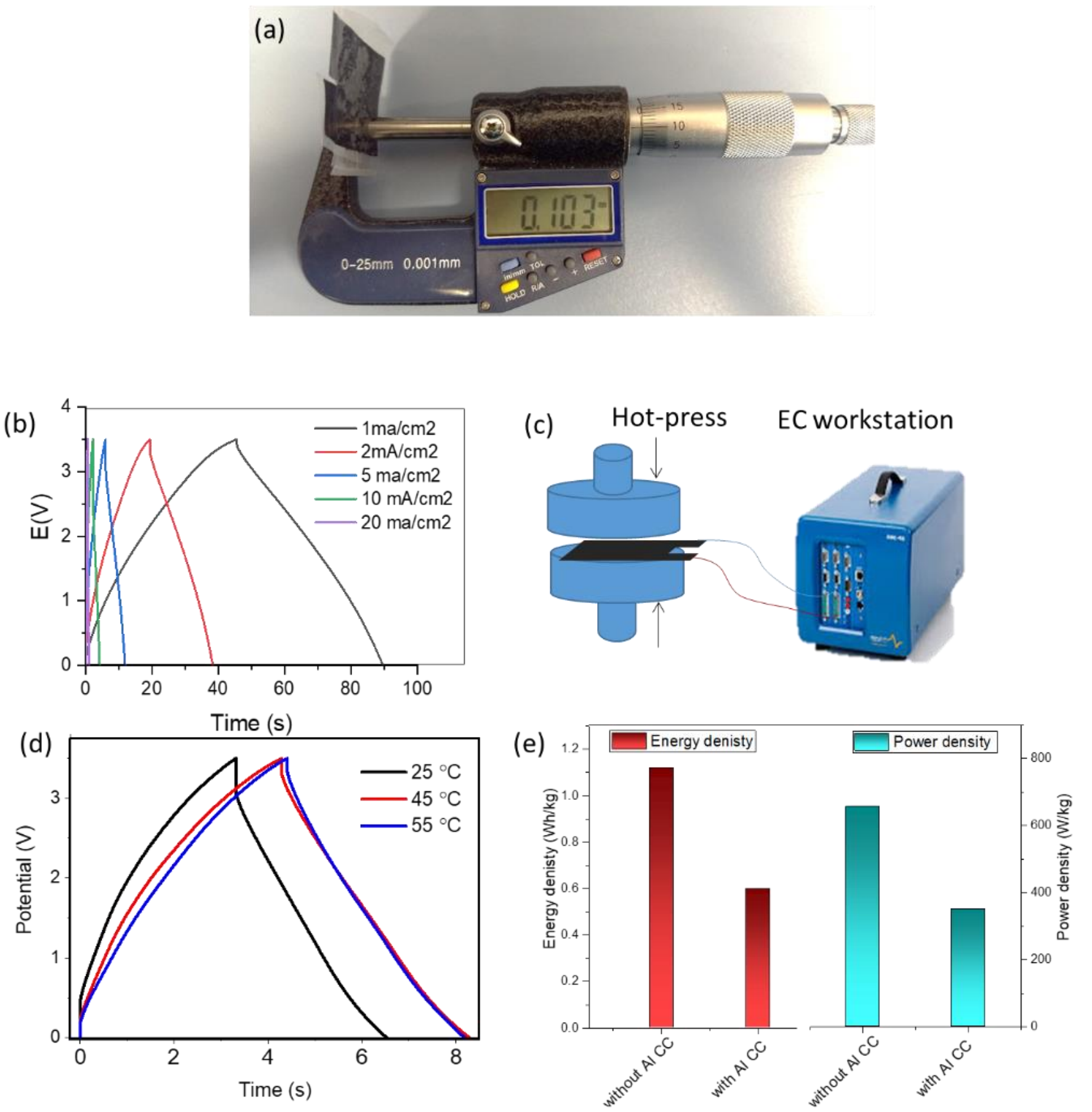

***Figure S1.*** *(a) Digital image of a free-standing EDLC during thickness measurement using a screw gage. (b) Charge-discharge profiles of the CNT-IL symmetric, free-standing device at different current densities. (c) Schematic of instrumental set up for in-situ thermal stability test of EDLC. (d) CD profiles of a free standing device at different temperatures. (e) Comparison of Energy density and power density of the EDLC with and without Al current collector.*

**Calculation of power and energy density of the composite:**

The real power (Pc) and energy density (Ec) of the composite were calculated using the following equations:

$$Pc = \frac{n \times Pcnt \times mass\ of\ CNT}{n \times mass\ of\ EDLC + dead\ weight\ of\ composite}$$

$$Ec = \frac{n \times Ecnt \times mass\ of\ CNT}{n \times mass\ of\ EDLC + dead\ weight\ of\ composite}$$

where *Pcnt* and *Ecnt* are the power and energy densities of CNT fiber fabrics, *n* (1-14) is number EDLC interleaves embedded in the composite.

Here the mass of EDLC can be expressed as following:

$$mass\ of\ EDLC = mass\ of\ (CNT + PEIL\ membrane + PVA\ membranes + Al\ current\ collectors)$$

For the devices without current collector, the mass of Al CC was nullified.

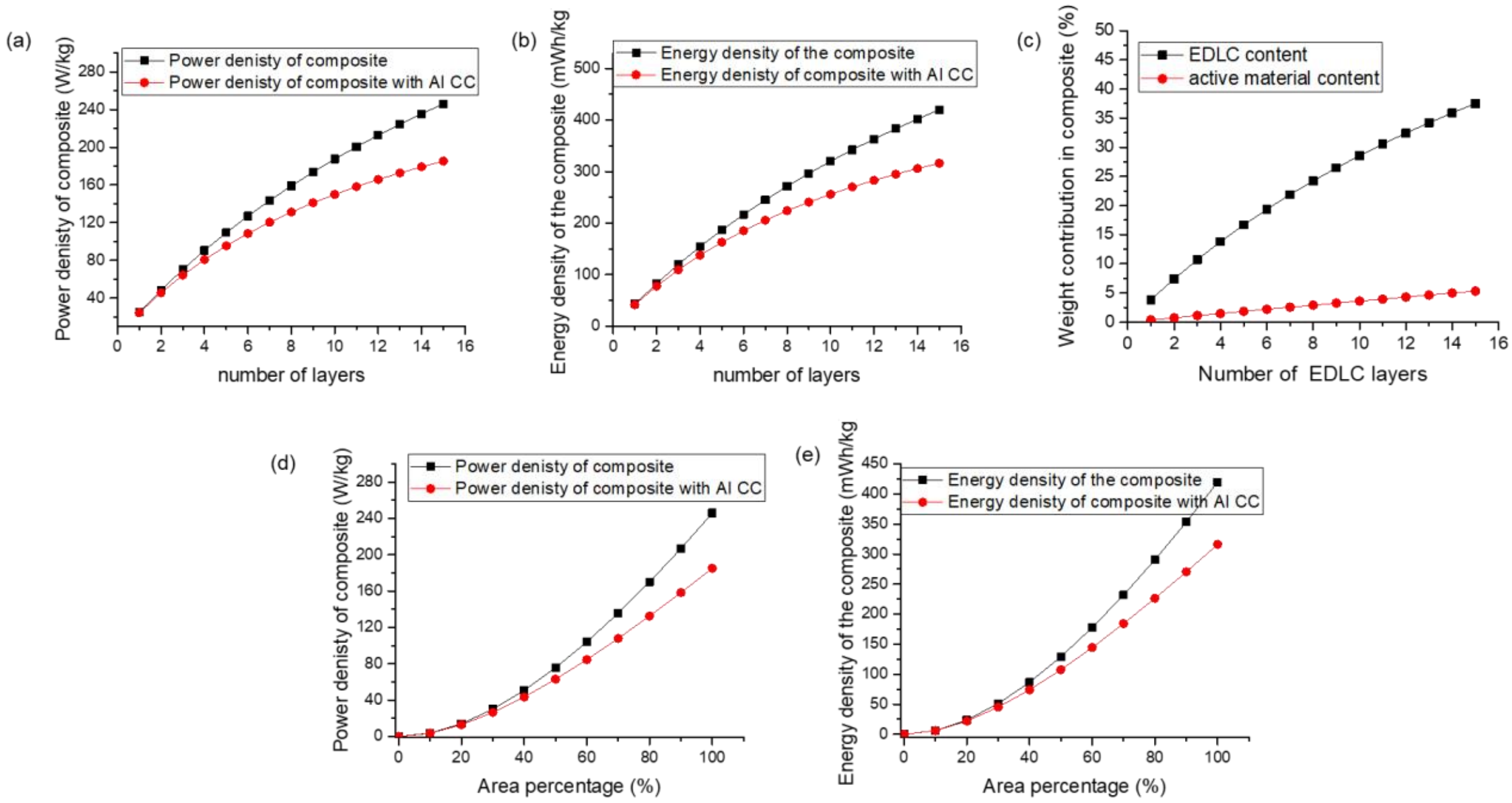

***Figure S2.*** *Variation of (a) power density and (b) energy density of the composite with respect to number of EDLC in the composite. (c) Weight contribution of the EDLC and active material (CNT) with increasing number of EDLC layers. Variation of (d) power density and (e) energy density of the composite with respect to effective area of 14 EDLC layers in the composite.*

**Single-lap shear Numerical simulation**

A finite element analysis (FEA) was performed with the commercial software Abaqus 2016. Explicit dynamic method was used to simulate the quasi-static loading case in order to conquer the convergence problem. The geometry corresponded to the dimensions presented in **Figure *S3*** with length, width and thickness of 200 mm, 48 and 3.1mm, respectively. Two CFRP adherends with a thickness of 1.5 mm were modelled with C3D8R elements, based on the properties listed in **Table ST1**. While the adhesive was modelled with cohesive element (COH3D8), using those values from **Table ST2**. The Supercapacitor was treated as a weak adhesive layer, which failed immediately when suffering shear/normal loads. Two notches were generated by assigning seams in Abaqus model. At the left side, all 6 Degrees of Freedom (DoF) are fixed, while on the right side, solely longitudinal displacement is allowed (x-direction) with a velocity of 0.15 mm/min.

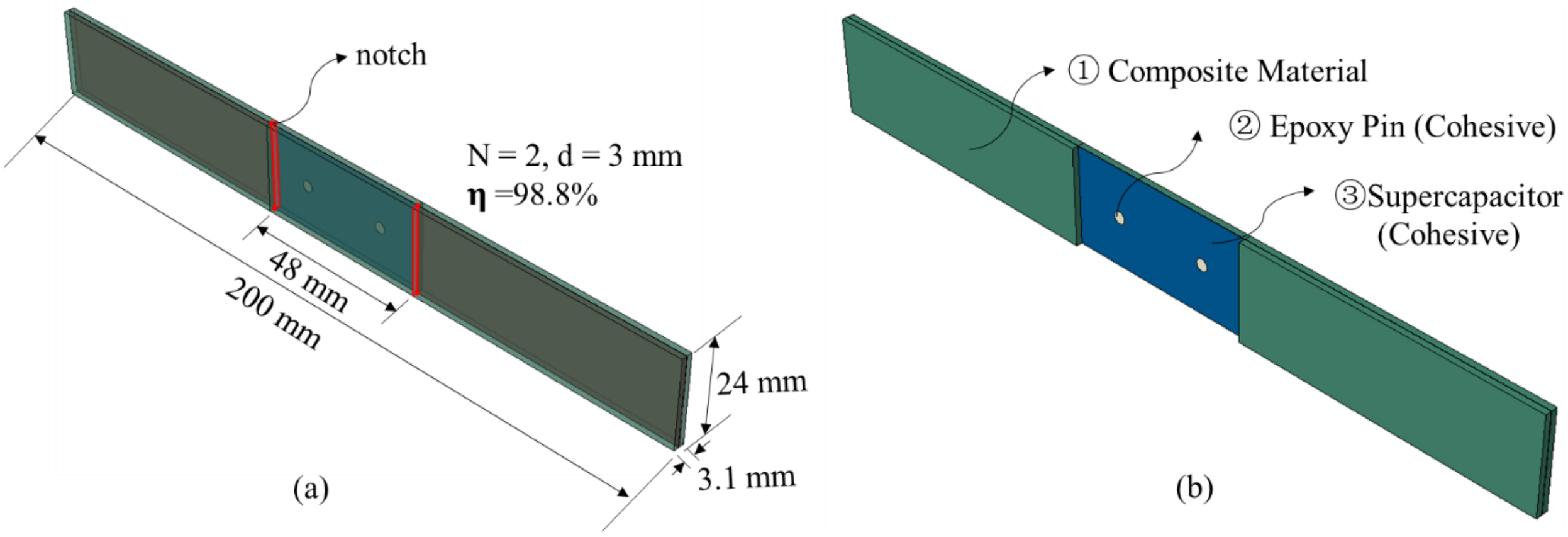

***Figure S3***. *(a) dimension of the single lap shear modeling and (b) material assignment*

***Table ST1.*** *Material properties of CFRP adherends*

| | | |
|---|---|---|
| Young's Modulus | E1 | 70e3 MPa |
| | E2 | 70e3 MPa |
| | *E3* | 7e3 MPa |
| Poisson ratio | Nu12 | 0.3 |
| | Nu13 | 0.3 |
| | Nu23 | 0.3 |
| Shear modulus | G12 | 5.5e3 MPa |
| | G13 | 5.5e3 MPa |
| | G23 | 5.5e3 MPa |

***Table ST2.*** *Material properties of epoxy film adhesive (Hysol® EA 9695™)*[1]

| | | |
|---|---|---|
| Tensile strength | $T_{Adh}$ | 48 MPa |
| Maximum elongation at break | $\varepsilon_{tmax}$ | 11.5 % |
| Tensile modulus | $E_{Adh}$ | 2019 MPa |
| Poisson ratio | $\nu_{adh}$ | 0.34 |
| Critical strain energy release rate (Mode I) | $G_{IC}$ | 1.019 kJ/m$^2$ |
| Critical strain energy release rate (Mode II) | $G_{IIC}$ | 0.783 kJ/m$^2$ |

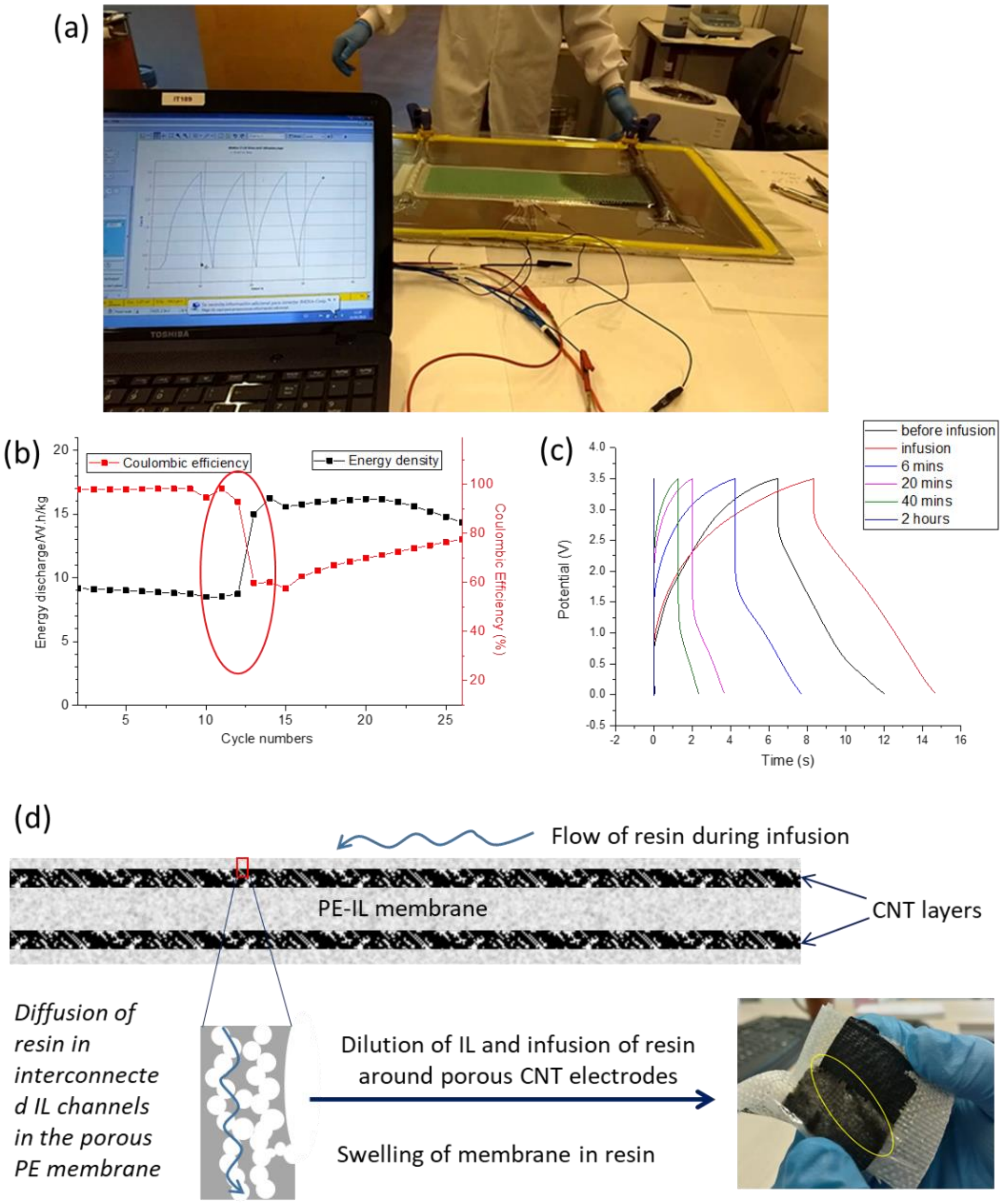

***Figure S4.*** *(a) Digital image of the experimental set up used during in situ electrochemical test during resin infusion process. (b) Plot of Energy and coulombic efficiency over CD cycles during the in-situ infusion experiment, where a free-standing device without PVA membrane was embedded in glass fibers. (c) Charge-discharge profiles of the device at different time after the infusion. (d) Schematic representation of the device failure mechanism showing percolation of liquid resin through the polymer electrolyte membrane during the infusion process, which leads to dilution of the electrolyte, followed by solidification of the membrane.*

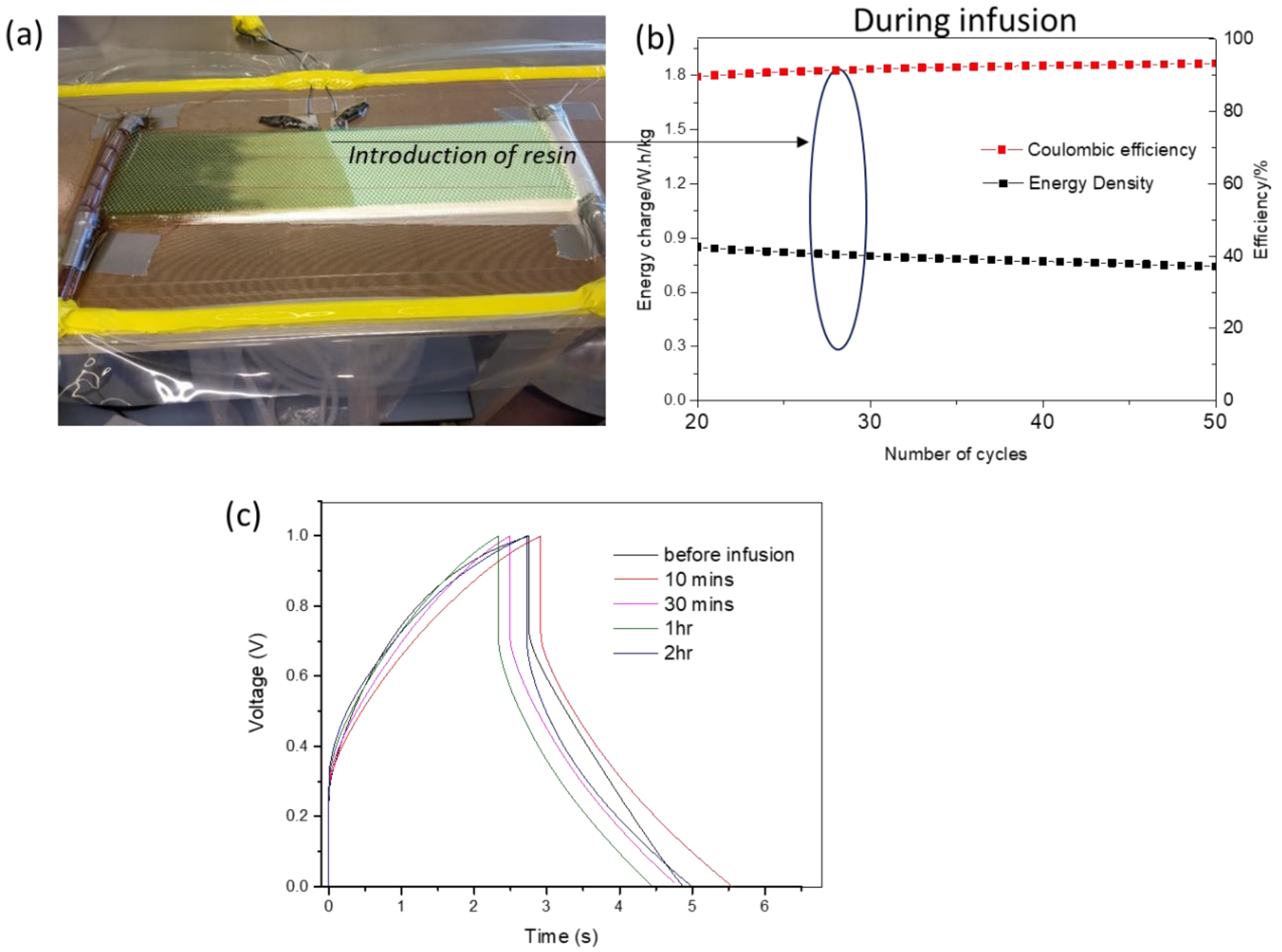

***Figure S5.*** *(a) Digital image of in situ infusion process, where a free-standing symmetric supercapacitor with PVA-$Na_2SO_4$ membrane is embedded inside glass fiber layers. (b) Variation of Energy density and Coulombic efficiency of the device during infusion process. (c) Charge-discharge profiles of the supercapacitor before and after infusion process.*

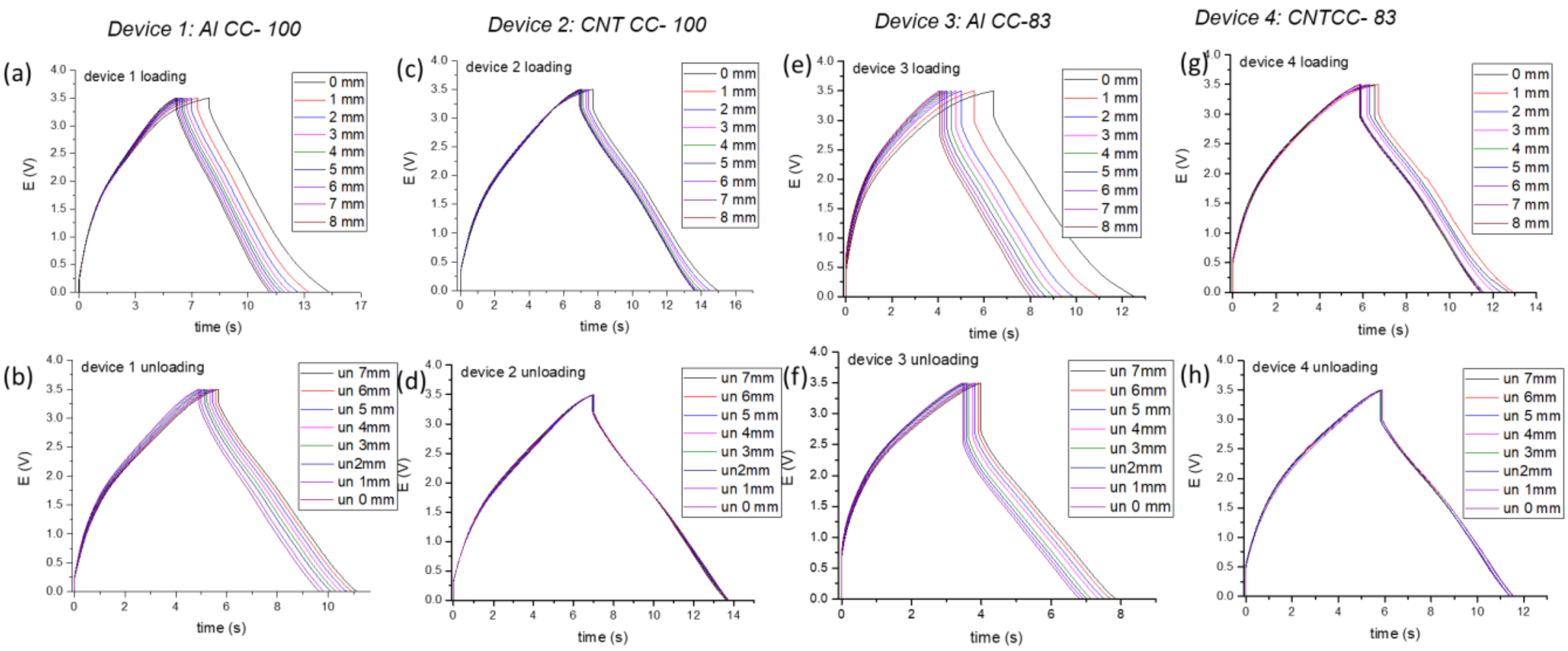

***Figure S6.*** *Charge-discharge (CD) profiles of the EDLC during step by step loading (a, c, e, h) and unloading (b, d, f, h) in the three-point bending tests for the non-structural (a-d) and structural devices (e-h) with (a, b, e, f) and without current collector (c, d, g, h).*

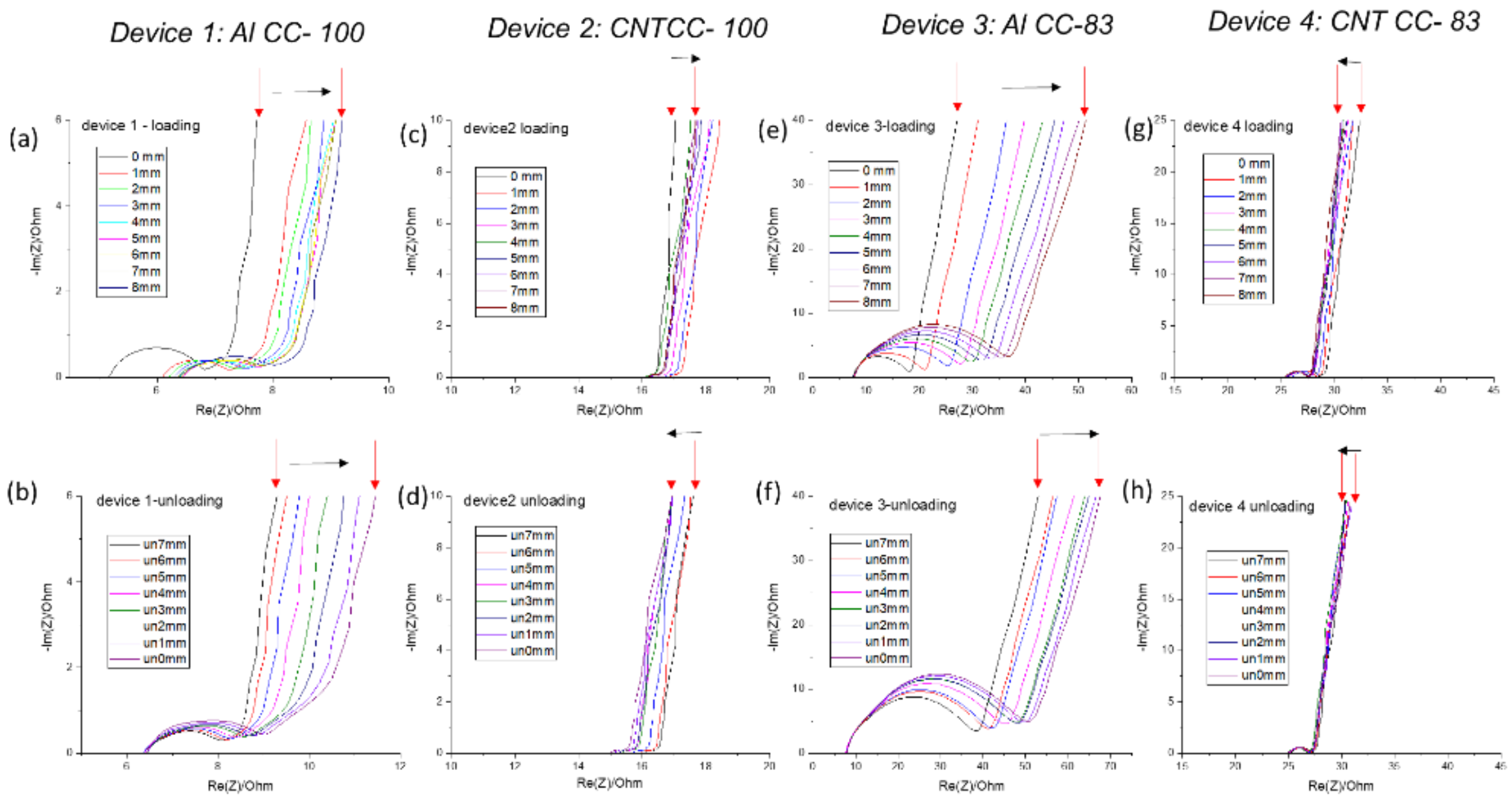

***Figure S7.*** *EIS spectra of the EDLC during step by step loading (a, c, e, h) and unloading (b, d, f, h) in the three-point bending tests for the non-structural (a-d) and structural devices (e-h) with (a, b, e, f) and without current collector (c, d, g, h).*

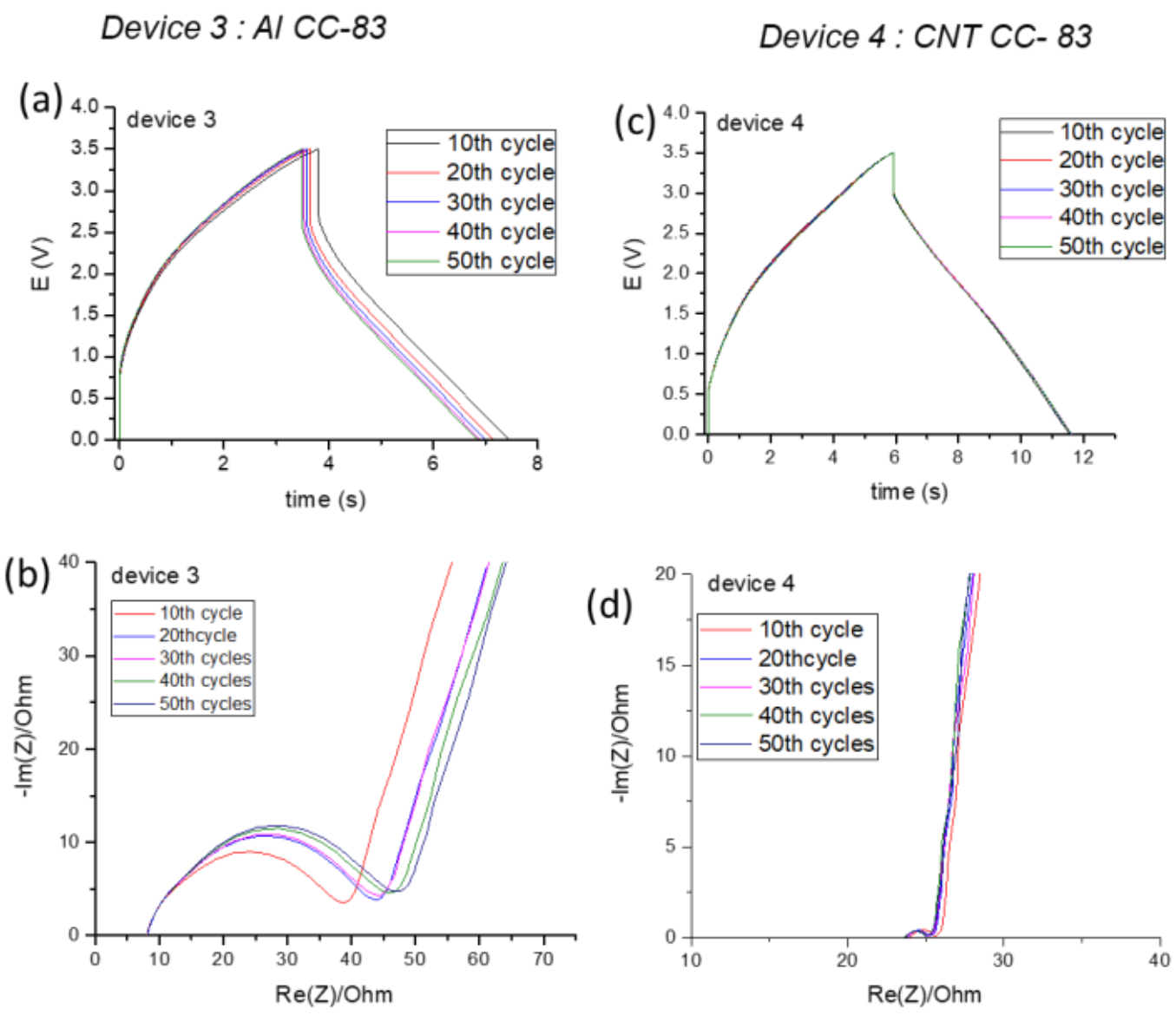

***Figure S8.*** *(a, b) charge-discharge profiles and (c, d) EIS spectra of the devices during cyclic fatigue test.*

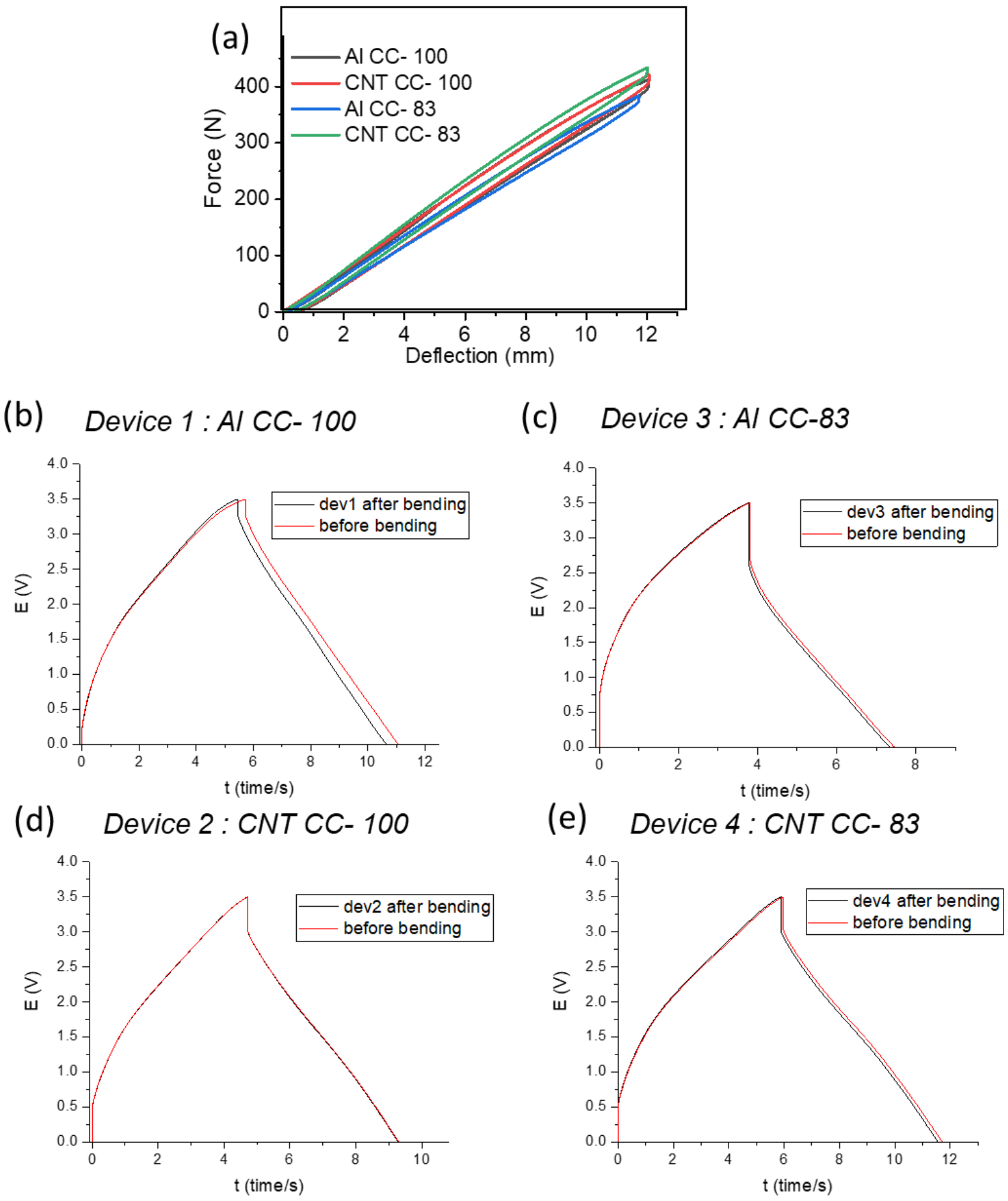

***Figure S9.*** *(a) Plot of force over deflection distance for the extreme bending test. (Comparison of charge discharge profiles of the EDLCs before and after extreme bending experiments (deflection length 12 mm).*

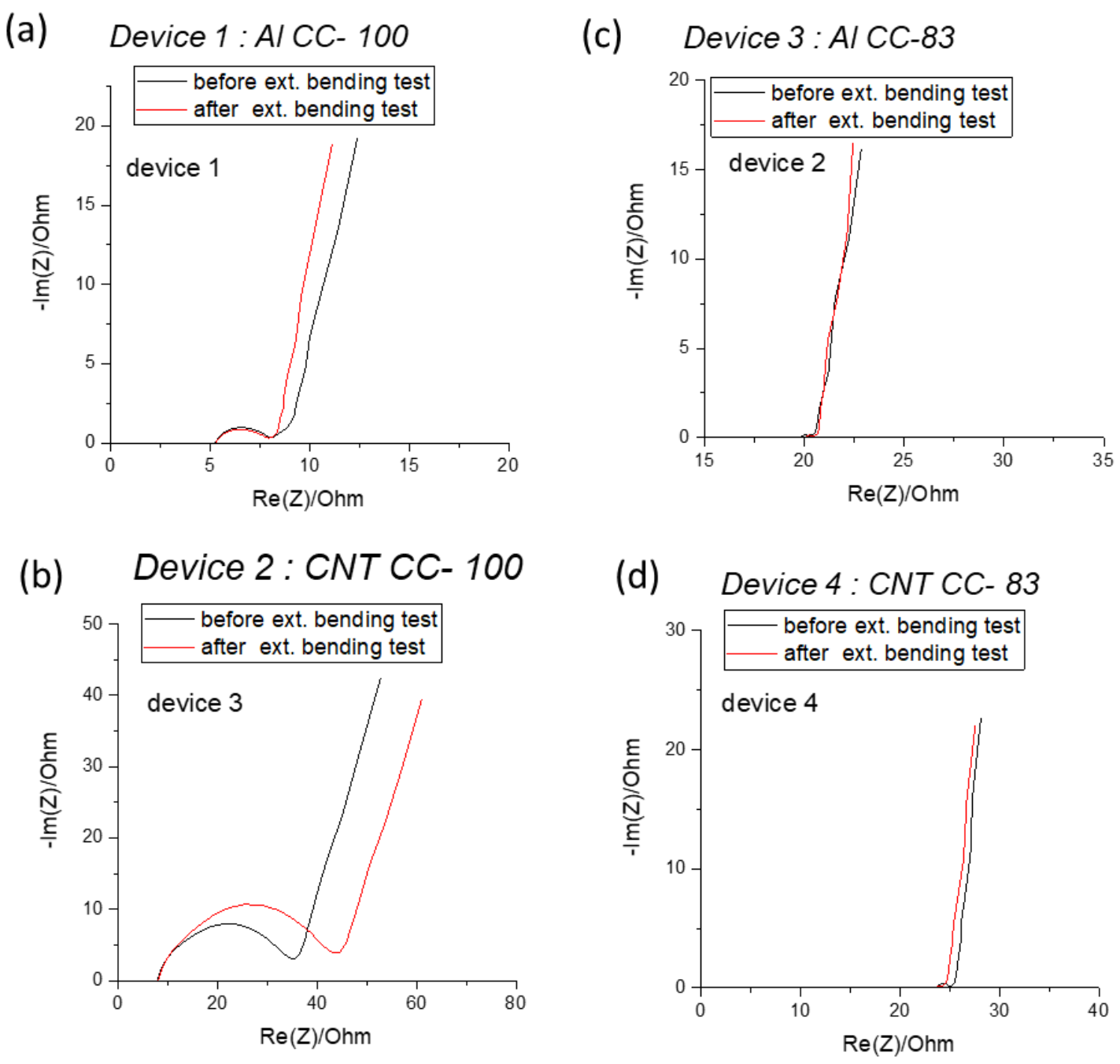

***Figure S10.*** *Comparison of EIS of the EDLCs before and after extreme bending experiments (deflection length 12 mm).*

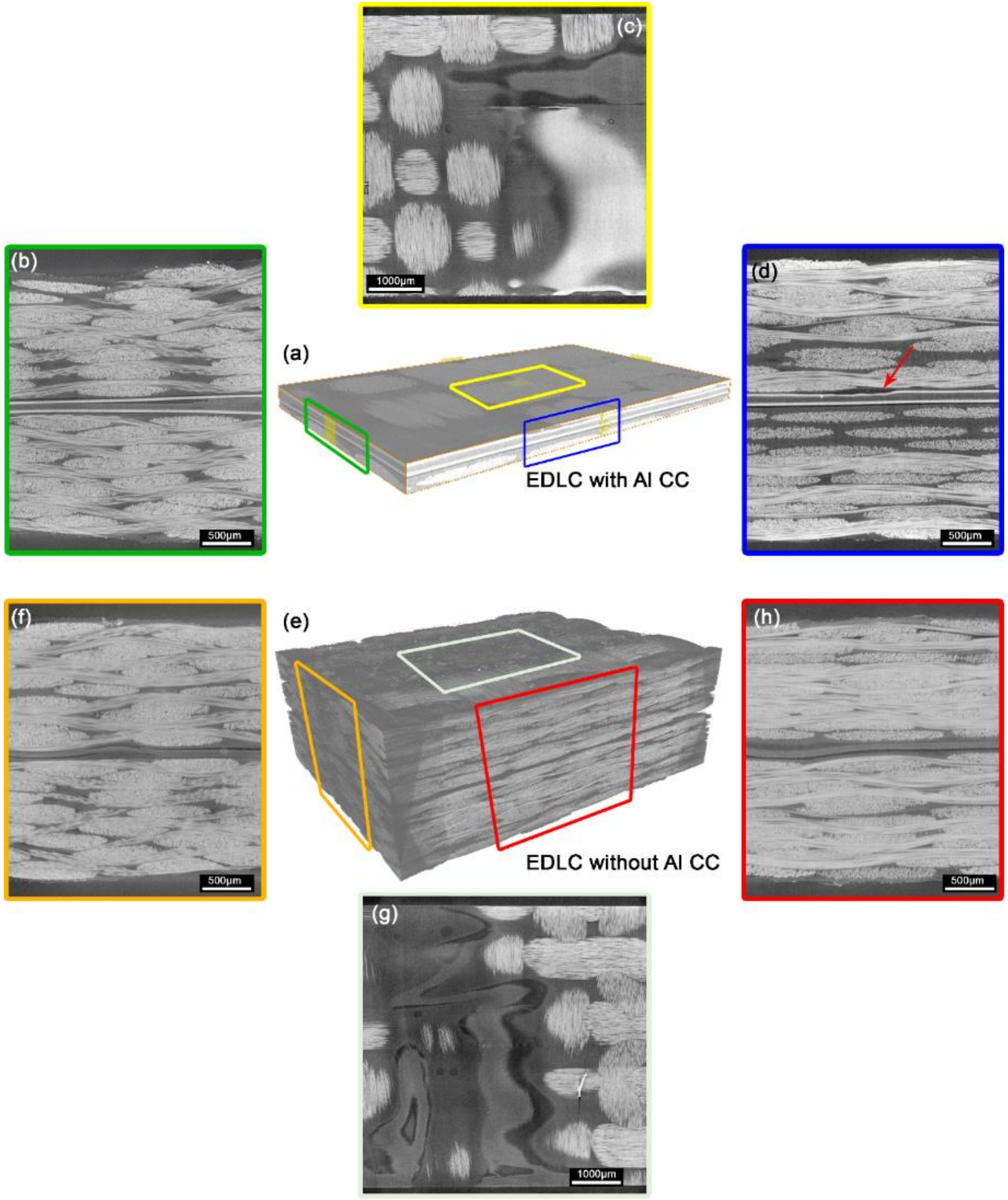

***Figure S11.*** *Results of 3D X-ray tomography experiments. Reconstructed image of a composite containing EDLC (a) with Al CC and (e) without CC from 3D X-ray tomography experiments. (b-d, f-g) Representative cross sectional images of the area around EDLC. The red arrow in (d) shows buckling of CC due to bending.*

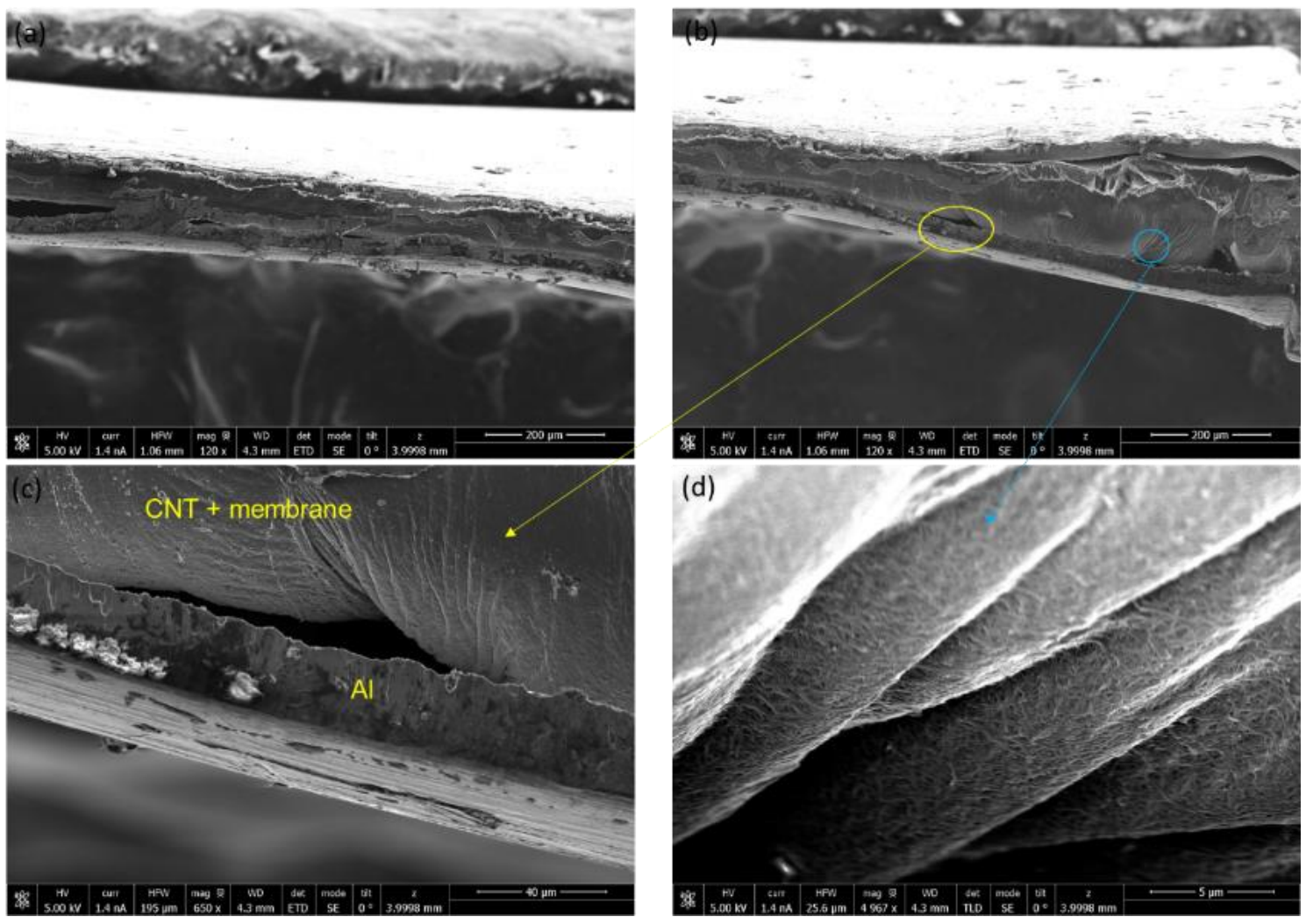

***Figure S12.*** *FESEM images of the EDLC with Al current collector after three-point bending experiments. (c) and (d) are the high resolution FESEM images around the area of detachment (Al/CNT-membrane) and CNT fabric embedded in polymer electrolyte membrane respectively.*

***Table ST3: Table to compare electrochemical and mechanical performance of structural charge storage devices***

| **Sr. no** | **Laminate component** | **Composite mass** | **Voltage window** | **Energy / Energy density** | **Power/ Power Density** | **Mechanical Properties** | **Reference** |
|---|---|---|---|---|---|---|---|
| 1. | Interleaves with CNT fiber veils PVDF-Pyr TFSI membrane, glass fiber and vinyl epoxy resin | 0.72 kg | 30 V (~3.5 V each) | 13.6 mWh | 7.97 W | Bending modulus (calculated) = 10 MPa | *This work* |
| 2. | NMC@Al, Graphite@Cu, polyolefin separator, LiPF6 in EC/DMC/DEC CFRP, resin: Bisphenol A diglycidyl ether, triethylene tetramine | 1.8 kg | 3.8 V | 240 Wh | NA | Bending modulus (calculated) = 10.5 GPa | [2] |
| 3. | Carbon fiber, carbon aerogel, 1-ethyl-3-methylimidazolium bis(trifluoromethylsulfonyl) imide (10 wt%), resin: polyethylene glycol diglycidyl ether based epoxy, Tri-ethylene-tetramine | NA | 0.1V | 0.001 Wh/kg | 0.033 W/kg | G12=8710 s12 =895 | [3] |
| 4. | Woven carbon fiber/ZnO nanotube, glass fiber separator, 1-ethyl-3-methylimidazolium tetrafluoroborate, lithium trifluoromethanesulfonate, resin: polyester resin, methyl ethyl ketone peroxide | NA | -0.75 V to 1V | 156.2 mWh/kg | 19.87 W/kg | tensile strength (325 MPa) and modulus (21 GPa) | [4] |
| 5. | Zn/PZB-931/γ-MnO2@Al, polyethylene terephthalate (PET) film | 5.6 g | 3.7 V (multiple in series, ~ 1.3V each) | Capacity ¬ 125 mAh/g | NA | Flexible, shape adaptable | [5] |
| 6. | Graphene/Aramid Nanofiber Composite, | NA | 1 V | 29.16 Wh/kg | NA | tensile strength (100.6 MPa) | [6] |
| 7. | CNT@ stainless steel mesh, $LiBF_4$ in 1-butyl 3-methyl imidizolium tetrafluoroborate ion conducting epoxy electrolyte matrix, CCR epoxy resin, Kevlar or fiberglass mats | NA | 2V | 3 mWh/kg | 1W/kg | elastic modulus >5GPa, mechanical strength > 85 MPa | [7] |